\documentclass{ws-rv10x7}
\usepackage[square]{ws-rv-van} 
\usepackage{ws-rv-thm}     
\usepackage{subfigure}

\usepackage[percent]{overpic}

\newcommand{\tw}{t_\mathrm{w}}
\newcommand{\Tc}{T_\mathrm{c}}
\begin{document}
\chapter[Numerical results on RSB]{Numerical Simulations and Replica Symmetry  Breaking\label{chapter_Numerical}}

\author[V\'{\i}ctor Mart\'{\i}n-Mayor, Juan J. Ruiz-Lorenzo, Beatriz Seoane and Peter Young]{V\'{\i}ctor Mart\'{\i}n-Mayor$^{(a)}$,
  Juan J. Ruiz-Lorenzo$^{(b)}$, Beatriz Seoane$^{(a)}$ and A. Peter Young$^{(c)}$
}

\address{a: Departamento de F\'{\i}sica Te\'orica, Universidad Complutense de Madrid,\\ Madrid 28040, Spain\\
  b: Departamento de F\'{\i}sica and Instituto de Computaci\'on Cient\'{\i}fica de Extremadura (ICCAEx), Universidad de Extremadura, 06006 Badajoz, Spain \\
  c: Department of Physics, University of California, Santa Cruz, California 95064, USA
}

\begin{abstract}
  Use of dedicated computers in spin glass simulations allows one to equilibrate
  very large samples (of size as large as $L=32$) and to carry out
  \emph{computer experiments} that can be compared to (and analyzed in
  combination with) laboratory experiments on spin-glass samples. In the
  absence of a magnetic field, the most economic conclusion of the combined
  analysis of equilibrium and non-equilibrium simulations is
  that an RSB spin glass phase is present in three spatial
  dimensions. However, in the presence of a field, the lower critical dimension for the de Almeida-Thouless transition seems to be larger than three.
\end{abstract}

\body

\section{Introduction}\label{sec:Intro}

Equilibrium numerical simulations have been an important tool used by the
scientific community to decide the theoretical controversy regarding the main
features of the spin glass phase (SG) at low temperatures $T<T_\mathrm{c}$
($T_\mathrm{c}$ is the critical temperature).
On one side of
this polemic, Parisi's solution of the SG in the mean field
approximation~\cite{mezard:87} has evolved into the replica symmetry breaking
(RSB) theory~\cite{marinari:00} according to which the spin-glass phase has many pure states. It can be regarded as as a critical
phase for all $T<T_\mathrm{c}$ in which the surfaces of the magnetic domains are
space filling. On the other hand, according to the droplet
theory~\cite{mcmillan:84,bray:87,fisher:86,fisher:88} there are only two pure states (in zero field) and the surfaces of magnetic domains (droplets) have a fractal dimension less than the space dimension $D$. It corresponds to the Migdal-Kadanoff approximation~\cite{gardner:84}. 
There is also an intermediate picture~\cite{krzakala:00,palassini:00} called TNT for ``trivial-non trivial".

However, the emphasis has changed somewhat in recent times. Recent numerical work has
mostly focused in out-of-equilibrium simulations (a choice partly motivated by
the fact that experimental work in spin glasses is carried out under
non-equilibrium conditions). The
so-called static-dynamic equivalence allows one to quantitatively relate 
quantities computed in equilibrium with out-of-equilibrium analogues (see
e.g. Ref.~\cite{franz:98,janus:08b,belletti2009depth,janus:10b,wittmann:16,janus:17,janus:21}). Dedicated computers, see
Sec.~\ref{sect:dedicated}, have had an important role in this  shift of focus that
has allowed for a new level of collaboration between simulations and experiments
in spin-glass physics. Indeed, it has become possible nowadays to subject
experimental and numerical data to a parallel analysis (see
Refs.~\cite{zhai2019slowing,zhai:2020,zhai-janus:21} and
Sec.~\ref{sect:out-equilibrium} in this chapter). From this new perspective,
the situation about the droplet/RSB polemic takes a different light depending
on the presence (or absence) of an external magnetic field:
\begin{itemize}
\item In the absence of a magnetic field, the latest results, both in
  equilibrium (see Sec.~\ref{sect:equilibrium}) and out-of-equilibrium
  (Sec.~\ref{sect:out-equilibrium}), find an RSB phase for $T<T_\mathrm{c}$
  and space dimension $D\!=\!3$. The droplet scenario still remains as a logical
  possibility, but only if one is willing to accept that current simulations
  and experiments are \emph{entirely} carried out far from the asymptotic
  regime.\footnote{A difficulty common to both simulations and experiments is that
    new behavior might emerge for larger lattice sizes
    or larger spin-glass coherence lengths.}
\item In the presence of a magnetic field, however, finding a spin-glass phase
  in $D\!=\!3$ has turned out to be extremely difficult. There is, however, evidence for a spin glass phase in a field in large $D$, see Sec.~\ref{sec:magnetic-field}.
\end{itemize}

Due to space limitations, we shall restrict ourselves to the case of Ising
spins ($S_{\boldsymbol x}=\pm1$), even though Heisenberg and XY spin glass models are
also interesting. Their transition temperature is surprisingly low,
the asymptotic scaling behavior has probably not been obtained accurately, and
there is controversy as to whether there is a separate transition involving
chiralities, with Kawamura arguing in favor of a separate
transition~\cite{kawamura:92,kawamura:98,kawamura:01,kawamura:03,takumi:20},
and other authors
disagreeing~\cite{fernandez:09b,campos:06,lee:03}. Nevertheless, universality
arguments suggest that the unavoidable residual anisotropies in the spin
interactions cause the distinction between Ising and Heisenberg spin glasses
to be asymptotically irrelevant~\cite{bray:82,baityjesi:14}. 
The question is not yet clear. On the one hand, maybe due to the difficulties in probing the asymptotic scaling regime, critical exponents do not seem to match. For instance, the exponent $\gamma$ for the non-linear susceptibility is approximately $5.5$ from simulations for the Ising spin-glass universality class~\cite{janus:13}, while experiments on Heisenberg spin glasses get lower values around 2 and 3~\cite{bouchiat:86}. On the other hand,  the quantitative agreement in the non-equilibrium dynamics in the spin glass phase between experiments in CuMn samples and Ising-Edwards-Anderson simulations, see Refs.~\cite{zhai2019slowing,zhai:2020,zhai-janus:21} and Sec.~\ref{sect:out-equilibrium}, seems to support the
choice of modeling spin glasses with an Ising Hamiltonian.

The remaining part of this chapter is organized as follows. We recall the standard model of spin-glasses in
Sec.~\ref{subsect:model}, define the main observables considered in
equilibrium simulations in Sec.~\ref{subsect:obs-equilibrium}, and review
finite-size scaling methods in Sec.~\ref{subsect:FSS}. The need for dedicated
computers is explained in Sec.~\ref{sect:dedicated}.  Important results
obtained in equilibrium simulations in the absence of a magnetic field are
recalled in Sec.~\ref{sect:equilibrium}. Equilibrium simulations in a field
are reviewed in Sec.~\ref{sec:magnetic-field}. Next, out-of-equilibrium
simulations are reviewed in Sec.~\ref{sect:out-equilibrium}, and finally, our conclusions are summarized in Sec.~\ref{sect:conclusions}.

\subsection{The Edwards-Anderson Model and its gauge symmetry}\label{subsect:model}

We shall be considering two geometries,
namely (hyper) cubic lattices in $D$-dimensions and $1D$ models with long-range interactions. In the cubic
geometry, the spins lie on the nodes of a (hyper)cubic lattice, whose  linear size is denoted by $L$, so the number of
spins is $N=L^D$. Periodic boundary conditions are usually taken and
interactions are typically restricted to lattice nearest-neighbors. The models with long-range interactions are in one dimension and the strength of the interactions falls off as a power of the distance between the spins. Varying this power is argued to be equivalent (at least roughly) to varying the dimension of the short-range model.

For both geometries, we consider the Edwards-Anderson (EA) Hamiltonian:
\begin{equation}\label{eq:EAH}
\mathcal H = - \sum_{\langle \boldsymbol x, \boldsymbol y\rangle}
J_{\boldsymbol x\boldsymbol y} S_{\boldsymbol x}S_{\boldsymbol y}-
\sum_{\boldsymbol x} h_{\boldsymbol x} S_{\boldsymbol x},
\end{equation}
where $\langle \boldsymbol x, \boldsymbol y\rangle$  indicates that the sum is
taken over all pairs of interacting spins (e.g.~nearest-neighbors for
the typical cubic geometry). It would be natural to use a uniform, external
magnetic field $h_{\boldsymbol x}=h$, but the gauge symmetry (see
below) makes it advisable to retain a site-dependence for the magnetic
fields $h_{\boldsymbol x}$. Spin glass models have quenched disorder
(see e.g.~Ref.~\cite{parisi:94}) in which the coupling constants
$J_{\boldsymbol x\boldsymbol y}$ (and sometimes also the magnetic fields
$h_{\boldsymbol x}$) are randomly extracted from a probability
distribution and held fixed once and for all. We call a
particular realization of the
$\{J_{\boldsymbol x\boldsymbol y},h_{\boldsymbol x}\}$ a \emph{sample}. Thermal averages,
denoted by $\langle\ldots\rangle$, are first computed for every sample. The
subsequent average over samples of the thermal mean-values is denoted
by $[\langle\ldots\rangle]$.

The couplings in Eq.~\eqref{eq:EAH} are independent, identically distributed
random variables. The most popular choices for the probability distributions
in a cubic geometry are the bimodal distribution
(in which $J_{\boldsymbol x\boldsymbol y}$ is $\pm1$ with $50\%$ probability) and a
Gaussian distribution with zero mean and unit variance. In the case of the
long-range geometry, one usually takes a Gaussian distribution.

The crucial role of the $Z_2$ Gauge symmetry of the Hamiltonian~\eqref{eq:EAH}
was soon realized~\cite{toulouse:77}. If one chooses $\epsilon_{\boldsymbol x}=\pm 1$  randomly at every lattice site $\boldsymbol{x}$, the energy remains invariant under
the transformation
\begin{equation}\label{eq:Gauge-Transform}
  S_{\boldsymbol x}\rightarrow \epsilon_{\boldsymbol x} S_{\boldsymbol x}\,,\quad
    J_{\boldsymbol x\boldsymbol y}\rightarrow \epsilon_{\boldsymbol x}
  J_{\boldsymbol x\boldsymbol y} \epsilon_{\boldsymbol y}\,,\quad
  h_{\boldsymbol x}\rightarrow \epsilon_{\boldsymbol x} h_{\boldsymbol x}\,. 
\end{equation}
For all
the standard choices of coupling distributions, one finds that the original choice $\{J_{\boldsymbol x\boldsymbol y},h_{\boldsymbol x}\}$
and its gauge-transformed values $\{\epsilon_{\boldsymbol x}
  J_{\boldsymbol x\boldsymbol y} \epsilon_{\boldsymbol
    y}\,\epsilon_{\boldsymbol x} h_{\boldsymbol x}\}$ occur with the same probability. Therefore, the sample-average $[\langle\ldots\rangle]$ effectively averages over all possible choices for the gauge parameters 
  $\epsilon_{\boldsymbol x}=\pm 1$.  All quantities that we
  focus on
  are invariant under the gauge transformation~\eqref{eq:Gauge-Transform}, see
  Secs.~\ref{subsect:obs-equilibrium},~\ref{subsect:obs-off-equilibrium}.

  \subsection{Observables (equilibrium)}\label{subsect:obs-equilibrium}

  A key quantity in our discussion will be
  the total overlap per spin defined by:
  \begin{equation}
    q\equiv q_{1,2}=\frac{1}{N} \sum_{\boldsymbol x} S_{\boldsymbol x}^{(1)}  S_{\boldsymbol x}^{(2)}\,,
  \end{equation}
where $S^{(1)}$ and $S^{(2)}$ are two real replicas of the
system with the same disorder. Its associated probability density function (pdf) averaged over the disorder  can be written as
\begin{equation}\label{eq:Pq-def}
  P(q) = \left [\left\langle \delta\left(q-\frac{1}{N} \sum_{\boldsymbol x} S_{\boldsymbol x}^{(1)}  S_{\boldsymbol x}^{(2)}\right)
    \right\rangle\right ] \,.
\end{equation}

It will also be useful to define the link overlap by 
\begin{equation}
Q_\mathrm{link}=\frac{1}{N_{\text{link}}} \sum_{\langle \boldsymbol x, \boldsymbol y\rangle} S^{(1)}_{\boldsymbol x} S^{(1)}_{\boldsymbol y}  S^{(2)}_{\boldsymbol x} S^{(2)}_{\boldsymbol y} \,,
\end{equation}
where the sum extends over all pairs of interacting lattice-sites (also known
as \emph{links}), whose number is ${N_{\text{link}}}$. In a $D$-dimensional
cubic lattice with periodic boundary conditions ${N_{\text{link}}}=DN$, but
for a long-range model ${N_{\text{link}}}$ may be as large as $N(N-1)/2$.
In mean field theory the link overlap is  trivially related to the spin
overlap by $Q_\mathrm{link}=q^2$.

Both $q$ and $Q_{\text{link}}$ are adequate for a mean-field treatment, but to go beyond this limit also we
need to consider the crucial role of fluctuations, characterized by
correlation functions. In the presence of a magnetic field, where individual
spins have a non-zero average, several different correlation functions can be
defined~\cite{dealmeida:78,dedominicis:06}. Specializing here only to the most
divergent correlations~\cite{parisi:13,fernandez:22}, we define the
``replicon'' propagator in real and Fourier space by
\begin{equation}\label{eq:G-replicon}
G(\boldsymbol{r}) = \frac{1}{N} \sum_{{\boldsymbol x}} \left[ \, \left(\,\langle S_{\boldsymbol x} S_{{\boldsymbol x}+{\boldsymbol r}} \rangle - \langle
S_{\boldsymbol x} \rangle \langle S_{{\boldsymbol x}+{\boldsymbol r}}\rangle
\, \right)^2 \, \right]\,,\quad \hat G (\boldsymbol k)=\sum_{\boldsymbol r}\
\mathrm{e}^{-\mathrm{i}{\boldsymbol k}\cdot{\boldsymbol r}}\, G(\boldsymbol{r})\,.
\end{equation}
In particular, the spin-glass susceptibility is
\begin{equation}\label{eq:chi-SG}
  \chi_{SG}\equiv \hat G \big({\boldsymbol k}=(0,0,\ldots,0)\big) = \frac{1}{N} \sum_{{\boldsymbol x},{\boldsymbol y}} \left[ \, \left(\,\langle S_{\boldsymbol x} S_{\boldsymbol y} \rangle - \langle
      S_{\boldsymbol x} \rangle \langle S_{\boldsymbol y}\rangle \, \right)^2 \, \right]\,.
\end{equation}
In the absence of a field, $\langle S_{\boldsymbol x}\rangle=0$ for all
sites ${\boldsymbol x}$ and Eq.~\eqref{eq:chi-SG} simplifies to
$\chi_{SG}=N\big[\langle q^2\rangle\big]$.

A quantity related to $\chi_{SG}$ is the second-moment correlation length (see
e.g.~\cite{amit:05}) that in a (hyper)cubic lattice with periodic boundary
conditions is
\begin{equation}\label{eq:xi2-def}
\xi_2=\frac{1}{2\sin (\pi/L)}\sqrt{\frac{\chi_{SG}}{\hat G ({\boldsymbol k}_1)}-1}\,,
\end{equation}
where ${\boldsymbol k_1}$ is the minimal non-vanishing wavevector allowed by
the boundary conditions (namely, ${\boldsymbol k}_1=(2\pi/L,0,0,\ldots,0)$ and
permutations). Interestingly, $\xi_2$ was instrumental in showing that there is a second order
spin glass phase transition in zero field in space dimension $D\!=\!3$~\cite{palassini:99,ballesteros:00}. A different definition of the
correlation length, more appropriate for out-of-equilibrium simulations, will be
discussed in Sec.~\ref{subsect:obs-off-equilibrium}.

Let us conclude this paragraph by recalling a most important quantity, namely
the Edwards-Anderson order parameter ($q_\mathrm{EA}$), which is the maximum
overlap. For mean-field models (such as the Sherrington-Kirkpatrick (SK)
model),
$q_{\mathrm{EA}}$ is given by
\begin{equation}
q_\mathrm{EA}=\sum_{\boldsymbol x} \big [\langle S_{\boldsymbol
  x}\rangle_\alpha^2 \big ]\,.
\end{equation}
where $\langle (\cdots) \rangle_\alpha$ is the average constrained to the
state $\alpha$ and is independent of the choice of the
state. Unfortunately, the definition of a \emph{state} beyond the mean-field
approximation is quite subtle (see the chapter by Newman, Read and Stein in this volume).

\subsection{Finite Size Scaling}\label{subsect:FSS}
The theory of finite-size scaling, see e.g.~\cite{amit:05}, explains how
critical divergencies are rounded in a finite-system of linear size $L$. Let $T_\text{c}(h)$ be
  the critical temperature, in which we have allowed a dependency on the magnetic field $h$, and let $O$  be a quantity 
diverging in the thermodynamic limit as $\big[\langle O\rangle \big]\propto
1/(T-T_\text{c}(h))^{x_o}$ (we refer only to the dominant divergence, there
  might be subleading terms). If the space dimension $D$ is smaller
  than the upper critical dimension $D_u$ (the dimension above which the mean-field
  approximation gives  exact values for critical exponents) then, according to finite-size scaling,
  \begin{equation}
    \big[\langle O\rangle \big](L,T)=L^{x_o/\nu} f_O(L^{1/\nu} t\bigr) + \ldots,\quad t = \frac{T-T_\text{c}(h)}{T_\text{c}(h)}\,,
  \end{equation}
  where $\nu$ is the thermal critical exponent and the dots stand for
  subleading scaling corrections.

  Of particular importance in this context are dimensionless quantities such
  as the second-moment correlation length $\xi_2$, defined in Eq.~\eqref{eq:xi2-def}, in
  units of the system-size  
  \begin{equation}\label{eq:xi-FSS}
\xi_2/L = f_\xi\bigl(L^{1/\nu} t\bigr) + \ldots \,.
\end{equation}
Dimensionless quantities are extremely useful to locate the critical point
see, for instance, Ref.~\cite{ballesteros:00}. An example of this use of
Eq.~\eqref{eq:xi-FSS} is
explained in Sec.~\ref{sec:magnetic-field} and Fig.~\ref{figH:xiLoverLYK2004}.

Some authors, working with $1D$ models with long-range interactions, have
found that, in the presence of an external magnetic field, the propagator
behaves anomalously, but only for the $\boldsymbol k=0$ mode
\cite{leuzzi:09}. This observation suggest trading $\xi_2/L$ for another
universal, renormalization-group invariant quantity named
$R_{12}$~\cite{janus:12}, defined by
\begin{equation}
R_{12} = \frac{\hat G (\boldsymbol k_1)}{\hat G(\boldsymbol k_2)},
\label{eq:R12}
\end{equation}
where ${\boldsymbol k_1}$ and ${\boldsymbol k_2}$ are the smallest
non-zero momenta compatible with the periodic boundary conditions.
For example, for $D\!=\!4$, $\boldsymbol k_1\!=\!(2\pi/L,0,0,0)$ and 
$\boldsymbol k_2 \!=\! (2\pi/L, \pm 2\pi/L, 0,0)$ (and permutations). The
generalization to other space dimensions is trivial.

\section{Why it is  so difficult to simulate spin glasses? The role of  dedicated computers.}\label{sect:dedicated}

Numerical simulation of spin glasses in equilibrium entails two major difficulties:
\begin{enumerate}
\item The variability between different samples is quite significant (see, for
  instance, sec.~\ref{subsect:replica-equiv}), which means that one needs to
  simulate a large number of samples in order to obtain an accurate sample
  average.
\item The simulation time need to equilibrate each sample is very
  significant. This is to be expected at zero temperature, because finding the
  ground gtate is an NP-complete problem~\cite{barahona:82b}. The problem
  remains difficult at finite temperatures. The most efficient Monte Carlo algorithm for spin glasses seems to be parallel tempering\cite{hukushima:96,marinari:98b}, also called ``replica exchange Monte Carlo''. Unfortunately there is no known, highly-efficient cluster algorithm of general applicability to spin glasses which corresponds to Swendsen and Wang's~\cite{swendsen:87} cluster algorithm for unfrustrated systems. In separate work, Swendsen and Wang~\cite{swendsen:86} developed a cluster-replica approach to spin glasses which works well in two dimensions~\cite{wang:05}. In higher dimensions, though, this approach effectively becomes equivalent to parallel tempering. A cluster, replica algorithm for $D\!=\!2$ spin glasses has also been developed by Houdayer~\cite{houdayer:01}. Since $T_c=0$ for two-dimensional spin glasses, so there is no low-temperature phase, the most efficient algorithm for the spin glass state below $T_c$ seems to be parallel tempering, as noted above.
  Even with the help of parallel tempering, some
  samples need an inordinately large equilibration time~\cite{fernandez:09b}. The
  underlying physical mechanism that hampers equilibration, even when using parallel tempering, seems to be
  temperature chaos~\cite{fernandez:13,billoire:18}.\footnote{It is remarkable
    that temperature chaos seems as well to be a major limiting factor for the
    performance a quantum annealer~\cite{martin-mayor:15}.}
\end{enumerate}

For out-equilibrium simulations, one uses very large samples to ensure that the slowly
growing, time-dependent coherence length $\xi(t)$ is much less than the system size $L$. One then needs fewer samples than for equilibrium simulations because one can
think  of a macroscopic sample as being
composed of $(L/\xi(t))^D$ equilibrated regions which are more or less independent of each other, so one large sample effectively averages over $(L/\xi(t))^D$ ``samples'' of size $L_\text{eff}\approx \xi(t)$. If one could simulate a truly macroscopic system then only one sample would be needed. This is, of course, the experimental situation.

Another difficulty in out-of-equilibrium simulations is that one has to mimic
natural dynamics using, for instance, the Metropolis algorithm. One is not allowed to use accelerated dynamics like parallel tempering. Unfortunately, natural dynamics is very slow at and below $T_\mathrm{c}$ (see e.g.~Fig.~\ref{fig:simulationsmeetexperiments}).
Hence, it is clear
that one needs to carry out very long simulations in order to reach reasonably
large values of $\xi(t)$. This topic is further elaborated in
Sec.~\ref{sect:out-equilibrium}.

Given these difficulties, a possible way forward is to build computers
specifically designed for spin-glass simulations. Several such computers have
been built over the years, such as the Ogielski machine~\cite{ogielski:85},
SUE~\cite{cruz:01}, and the Janus supercomputers~\cite{janus:08,janus:14}. The Janus II is currently the most powerful computer for spin glass simulations.

\section{Equilibrium numerical studies of the overlap}\label{sect:equilibrium}

RSB makes many predictions regarding the order
parameter in spin glasses. In this section we will focus
on a small number of them: (i) its density
probability function $P(q)$, (ii) overlap equivalence, i.e.~all possible
definitions of an overlap in the model encode the same physics and  (iii) stochastic stability (which is used to show the
Guerra's relations) and  ultrametricity.

\subsection{Structure of the Equilibrium $P(q)$}\label{seq:Pq}
\begin{figure}[t]\centering
\begin{overpic}[width=5.in,trim=30 140 30 155 ,clip]{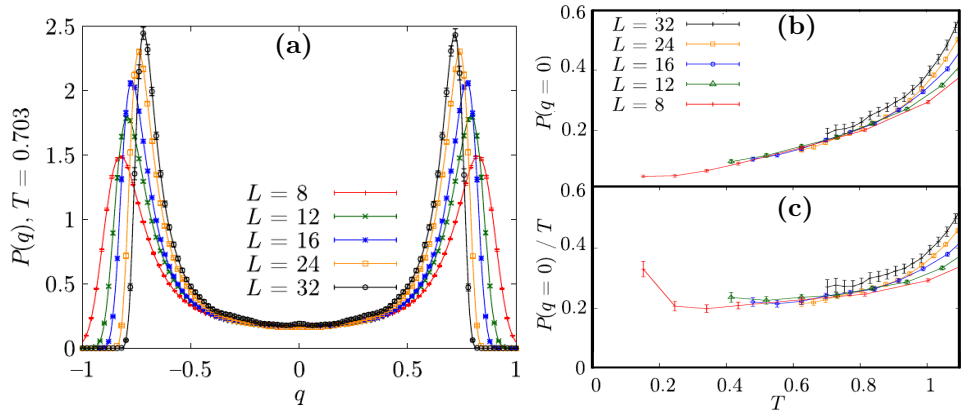}
\put(28,38){{\bf (a)}}
\put(80,40){{\bf (b)}}
\put(80,21){{\bf (c)}}
\end{overpic}
\caption{(\textbf{a}) Overlap density distribution function $P(q)$, see Eq.~\eqref{eq:Pq-def}, as computed for the $3D$ EA model with binary couplings at $T=0.703\!\approx\! 0.64T_{\mathrm{c}}$ for different lattice sizes $L$. (\textbf{b}) is $P(q=0)$ and (\textbf{c}) is $P(q=0)/T$ versus temperature $T$, as computed in $3D$ for several system sizes. We observe an envelope curve with a linear behavior, 
as expected from RSB. Figure adapted from\cite{janus:10}.}\label{fig:Pq}
\end{figure}
One manifestation of the infinite number of
pure states predicted by RSB theory 
is a non trivial pdf of the order parameter, $P(q)$, see Fig.~\ref{fig:Pq}(a). In 
contrast, the droplet model predicts a
trivial $P(q)$ in the thermodynamic limit, in the sense that it consists, for $h=0$, of  two Dirac-delta functions at $q\! =\! \pm q_\mathrm{EA}$, where $q_\mathrm{EA}$ is the Edwards-Anderson overlap. Instead, in RSB,
$P(q)$ has, in addition to these two delta functions, a continuous function in between which is non-zero in the thermodynamic limit.
In the droplet picture the continuous part of the distribution vanishes slowly with linear system size $L$ like $L^{-\theta_S}$ where $\theta_S$ is a stiffness exponent whose value is around $0.24$~\cite{boettcher:05} for $D\!=\!3$. 

Over the years there have been many studies~\cite{reger:90,berg:98,marinari:98d,katzgraber:01,katzgraber:02,berg:02,janus:10,wang:20} of the weight of $P(q)$ around $q\!=\!0$ and these consistently find a value independent, or nearly independent, of size, in agreement with RSB theory.

We show some recent results in $3D$~\cite{janus:10} for $P(q)$, Fig~\ref{eq:Pq-def}(a),
$P(0)$, Fig~\ref{eq:Pq-def}(b),
and $P(0)/T$,  Fig~\ref{eq:Pq-def}(c), as a function of the temperature deep in the spin
glass phase. We recall that $T_c=1.1019(29)$\cite{janus:13}. This data supports the
RSB predictions that $P(0)$ is independent of $L$, and is proportional to $T$ at low $T$. A different analysis of the $P(q)$
behavior comes to the same conclusions~\cite{wang:20}.
The behavior of $P(q\!=\!0)$ could be modified by the presence of
interfaces, so it was proposed to compute $P(q)$
in small boxes in order to avoid their effects.  This
analysis was performed in Ref.~\cite{marinari:98c}, which found the same behavior as
that obtained from the overlap computed over the whole lattice.

\subsection{Overlap Equivalence}

Overlap equivalence states that all possible definitions of new
overlaps in a spin glasses must be a function of the overlap $q$.
Equivalently, we can classify a pair of replicas using their overlap
and argue that no finer classification is possible (separability). This
implies that fluctuations of all reasonable definitions of the overlap will vanish if  
we work in a $q$-fixed ensemble.

The variance of a $q$-conditioned observable $O$ is defined by
\begin{equation}\label{eq:var-q}
\mathrm{Var}(O|q=c) = \mathrm{E}(O^2 | q=c) - \mathrm{E}(O|q=c)^2\,.
\end{equation}
where the symbol $\mathrm{E}(\cdot)$ denotes (i) the average in a
given sample over the configurations which satisfy the constraint ($q\!=\!c$), and
then (ii) the average over disorder.

\begin{figure}[t]
\centering
\includegraphics[trim=25 30 25 30, clip,height=0.55\linewidth,angle=270]{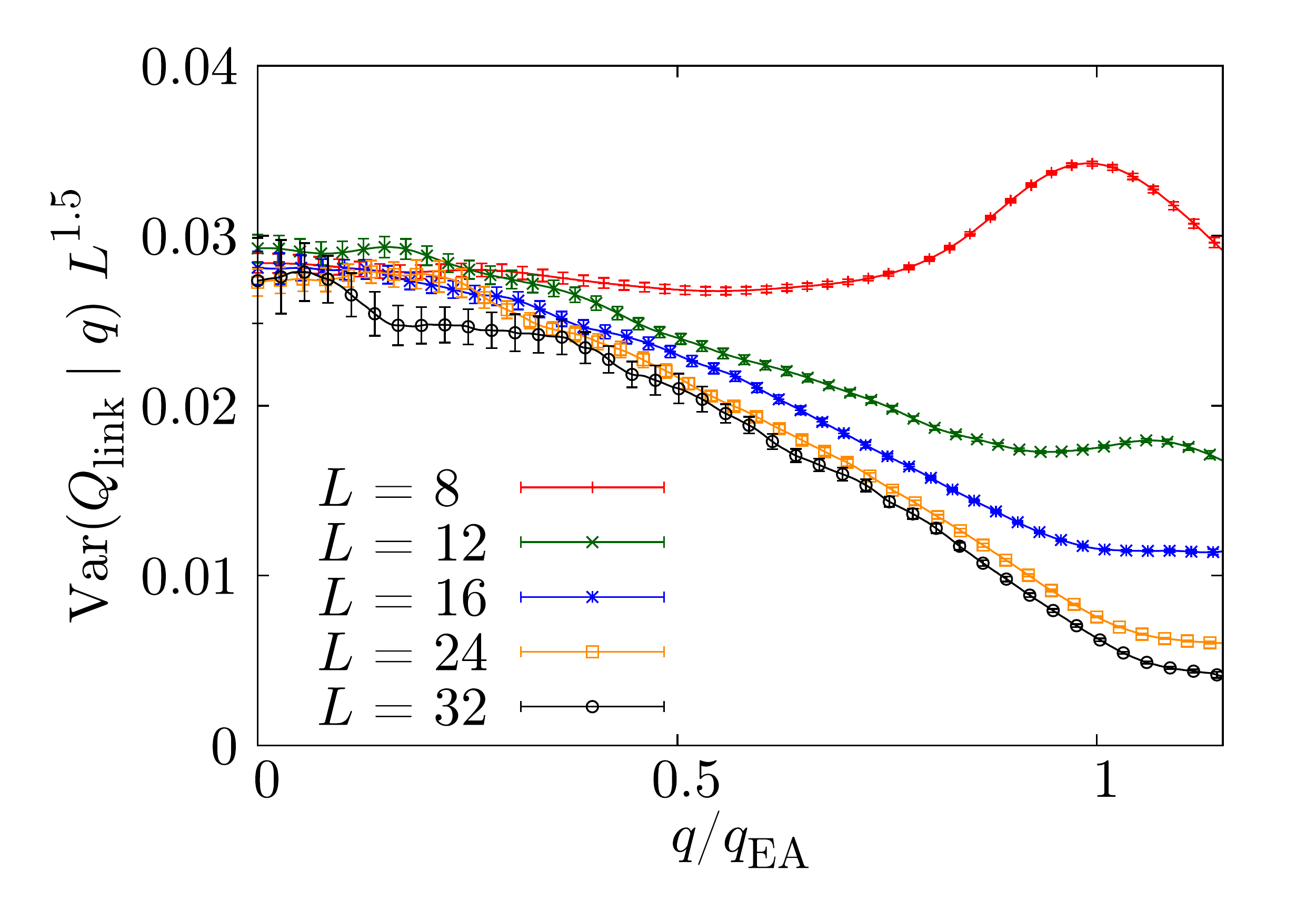}
\caption{Plot of the conditional variance at fixed $q$ of $Q_\mathrm{link}$, see Eq.~\eqref{eq:var-q},
rescaled by appropriate powers of $L$ in $3D$ at $T=0.703$ (we chose
exponents that provided a good scaling at $q\!=\!0$). The abscissa corresponds to
$q$ in units of $q_\mathrm{EA}(L,T=0.703)$. Figure taken from \cite{janus:10}.
\label{fig:var-qlink}
}
\end{figure}

In this subsection we address the behavior of
$\mathrm{Var}(Q_\mathrm{link} | q)$. RSB predicts that this conditioned variance should go to zero in
the limit of large lattices, due to the existence of a relation
between $Q_\mathrm{link}$ and $q$. In the droplet model the only possible value of $q$ is $q_\mathrm{EA}$, so $Q_\mathrm{link}$ will be defined only for this value of the overlap (in the infinite volume limit). Therefore, the droplet model predicts as well a vanishing conditioned variance. 

Indeed, Fig. \ref{fig:var-qlink} shows that
this conditioned variance goes to zero in the limit of large $L$\cite{janus:10}. Moreover, Ref. \cite{contucci:06} reaches the same conclusion from a different analysis.  The good scaling of the conditioned variance of $Q_\mathrm{link}$ near $q\!=\!0$  extends to $q_{EA}$ for the larger lattices ($L\!=\!24$ and $L\!=\!32$). This scaling supports RSB and not droplet, because in the droplet picture it would be natural to expect completely different finite size effects for $q\!=\!q_\mathrm{EA}$ and for $q<q_\mathrm{EA}$  (and in particular for $q\!=\!0$). 

Another interesting quantity is  $\mathrm{d} \mathrm{E}(Q_\mathrm{link}|q)/\mathrm{d} q^2$. In the droplet model, at variance with the RSB picture, this derivative should be zero. Numerically, the derivative is nonzero, although its size decreases with $L$. Therefore, the analysis based on this observable is not conclusive~\cite{janus:10}. Nevertheless the results of Ref.~\cite{contucci:06}, present numerical evidences  that $\mathrm{E}(Q_\mathrm{link}|q^2)$ is an increasing one-to-one function of $q^2$ in the thermodynamic limit. This fact, when combined with a vanishing conditioned variance of $Q_\mathrm{link}$, rules out the TNT description of the spin-glass phase.

\subsection{Stochastic Stability}\label{subsect:replica-equiv}

Stochastic stability\footnote{The Parisi matrix $Q_{ab}$ satisfies the condition that 
  $\sum_b f( Q_{ab} )$
is independent of the replica index $a$. Stochastic stability states
the invariance of the distribution of the free energies under
independent random increments of the interactions. Stochastic invariance implies replica invariance but is more general.} 
has been proved in the SK model by Guerra~\cite{guerra:96,ghirlanda:98} and by Aizenman and Contucci~\cite{aizenman:98}.  In this section we present numerical evidence supporting stochastic stability in finite dimensional models, which, in turn, provides evidence that RSB applies in those systems.

Using stochastic stability is possible  to write the following relation,
\begin{equation}\label{eq:R-link}
R_{\mathrm{link}}=\frac{\big [ \langle Q_\mathrm{link}^2 \rangle - \langle Q_\mathrm{link} \rangle^2\big ]}{\big [\langle Q_\mathrm{link}^2 \rangle \big ]\ -\ \big [ \langle Q_\mathrm{link} \rangle\big ]^2} = \frac{2}{3}\qquad (\mathrm{RSB\ }, L \to\infty, T < T_c)  \,.
\end{equation}
Note the difference in the placement of the square in the subtracted terms in the numerator and denominator.
In the droplet or TNT pictures both the numerator and denominator vanish in the thermodynamic limit.

Another analogous observable with the same RSB behavior is
$R_{q^2}$ (using the mean-field correspondence
$Q_\mathrm{link}\rightarrow q^2$):
\begin{equation}\label{eq:R-q2}
R_{q^2}=\frac{\big [ \langle q^4 \rangle\ -\ \langle q^2 \rangle^2 \big ]}
{\big [\langle q^4 \rangle \big ]\ -\ \big [\langle q^2 \rangle \big ]^2}
= \frac{2}{3}\qquad (\mathrm{RSB\ }, L \to\infty, T < T_c) \,.
\end{equation}

\begin{figure}[t]
\centering
\begin{minipage}{.48\linewidth}
\begin{overpic}[height=\linewidth,angle=270]{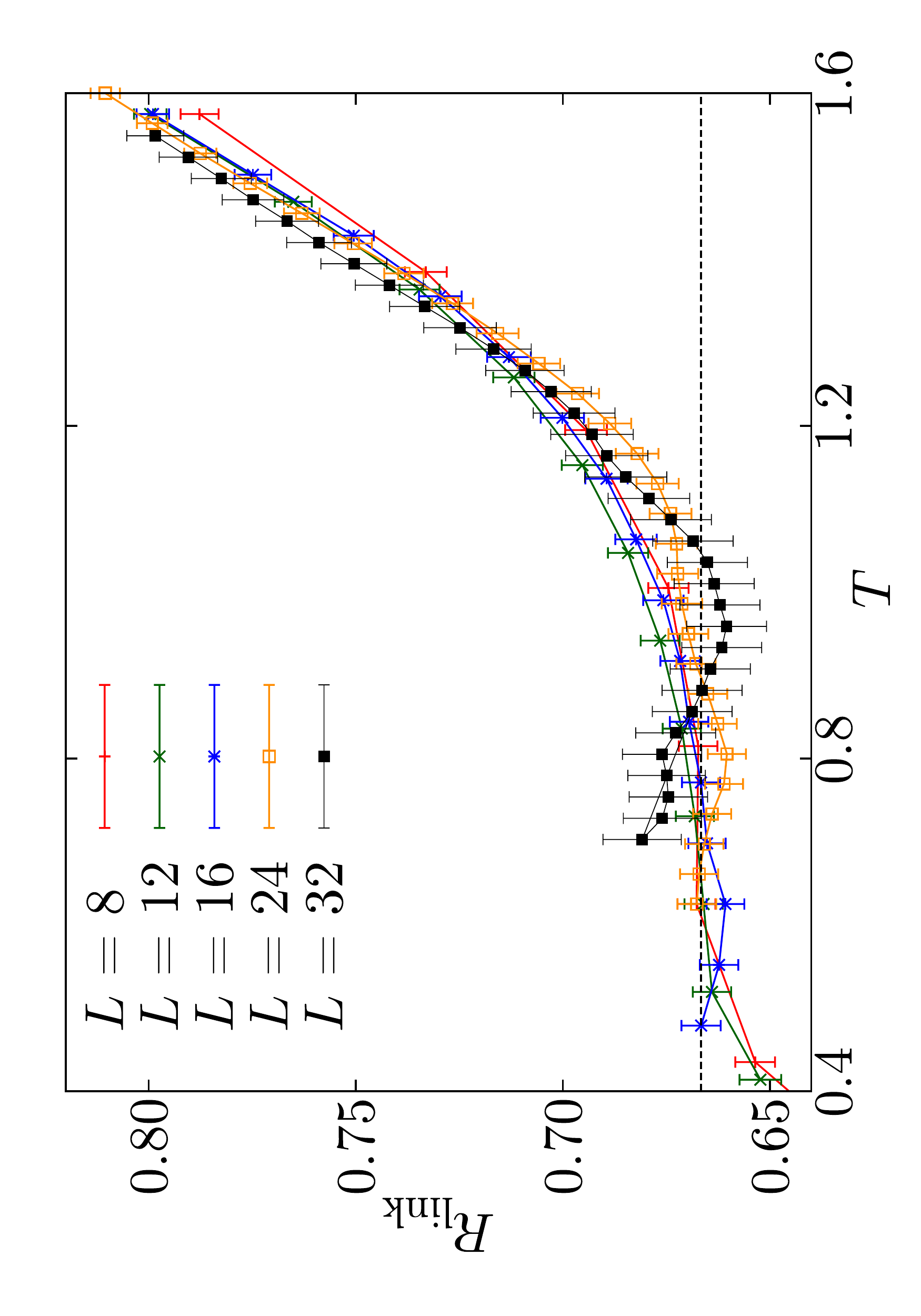}
\put(55,60){{\bf (a)}}
\end{overpic}
\end{minipage}
\begin{minipage}{.48\linewidth}
\begin{overpic}[height=\linewidth,angle=270]{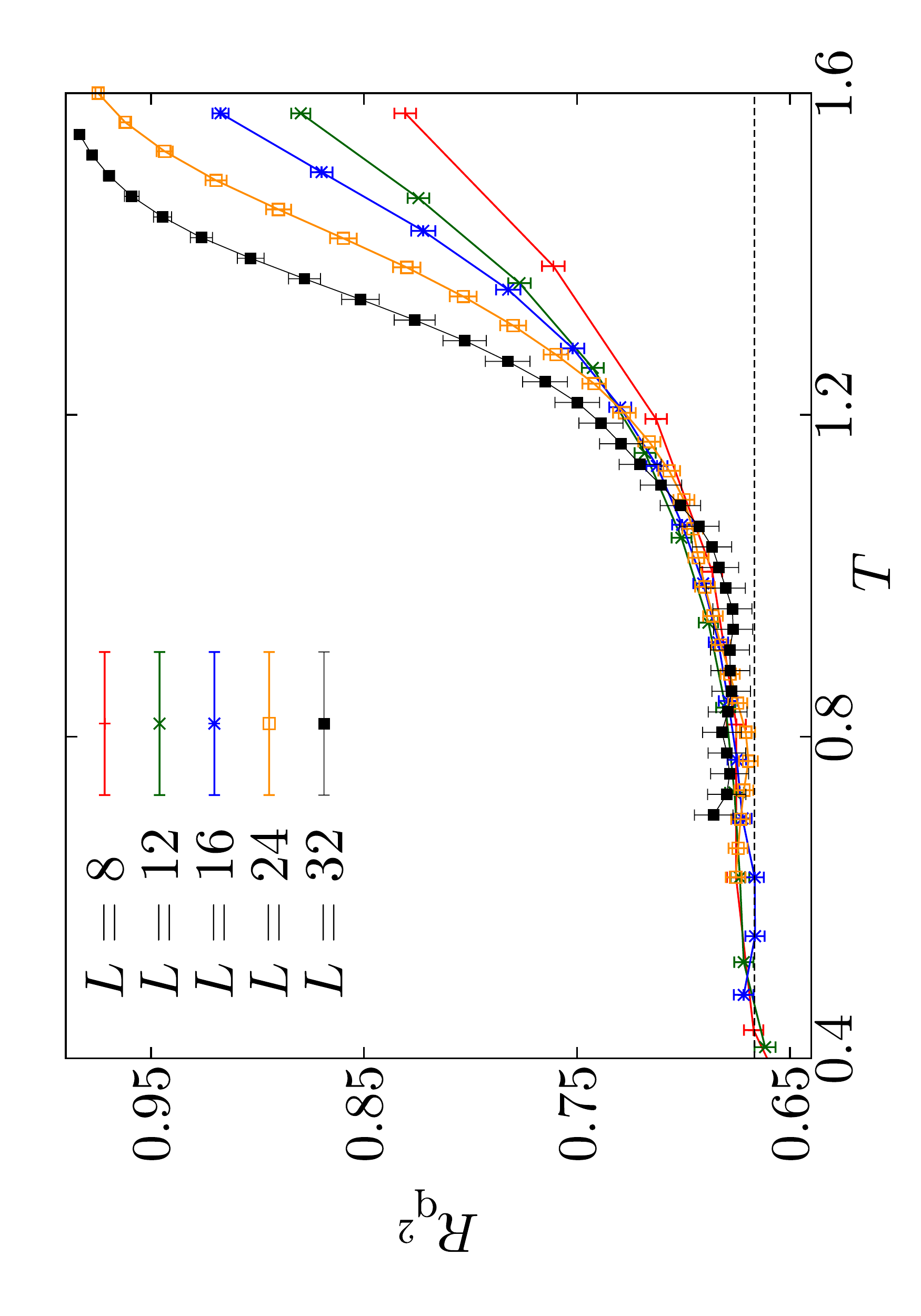}
\put(55,60){{\bf (b)}}
\end{overpic}
\end{minipage}
\caption{The ratios $R_\mathrm{link}$, Eq.~(\ref{eq:R-link}), (\textbf{a})
  and $R_{q^2}$, Eq.~(\ref{eq:R-q2}), (\textbf{b}) versus $T$
for the different system sizes in $3D$. Stochastic stability  implies
that, in an RSB system below $T_\mathrm{c}$, $R^\mathrm{link} =R_\mathrm{q^2} = 2/3$ 
in the large-$L$ limit. Recall that $T_\mathrm{c} \approx 1.1$. Figure taken from~\cite{janus:10}.}
\label{fig:estabilidad}
\end{figure}

In Fig. \ref{fig:estabilidad} we show the behavior of
$R_\mathrm{link}$ and $R_{q^2}$ as a function of temperature in $3D$ for different lattice sizes. The convergence of both observables to
the mean field limit, 2/3, is very good below the critical
temperature. For additional analysis, see Ref.~\cite{contucci:06}.

Finally, we briefly discuss the issue of ultrametricity.  It is possible to show that overlap equivalence and stochastic stability imply ultrametricity~\cite{parisi:00}. In addition, using some of the
Guerra's relations~\cite{guerra:96,ghirlanda:98} and  assuming the existence
of ultrametricity in finite space dimension $D$,  one finds that ultrametricity at finite $D$ should have the same properties that one finds for $D\!=\!\infty$ (i.e. a
quarter of the triangles are equilateral, otherwise they are
isosceles)~\cite{iniguez:96}. Furthermore, Panchenko has shown that ultrametricity follows from stochastic stability without additional assumptions~\cite{panchenko:13}.
Hence, the results already presented in this section indicate that ultrametricity should exist in three-dimensions with the same
properties as in mean field theory.  However, direct
detection of ultrametricity in 3$D$ spin glasses remains elusive~\cite{hed:04,janus:11}.

\section{Results in a magnetic field}\label{sec:magnetic-field}

\begin{figure}[t]
\centerline{\includegraphics[trim=0 0 0 0,clip, width=4.in]{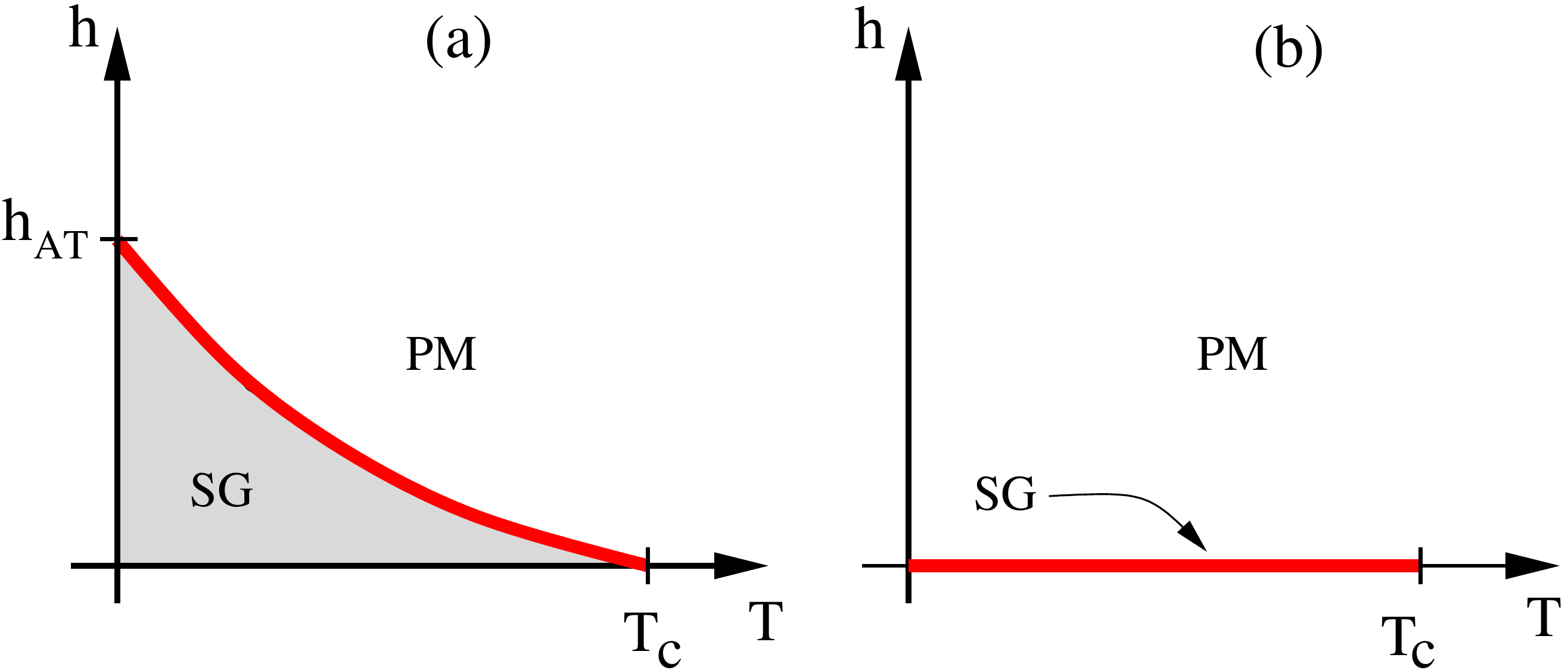}}
\caption{Schematic representation of the magnetic-field, temperature, ($h,T)$,
  phase diagram for a spin-glass. According to RSB theory, the phase diagram depends strongly on the
  space dimension.
 (\textbf{a}) If
  the space dimension is larger than the lower critical dimension
  (in a field) $D>D_{\mathrm{\ell}}^h$, there is a spin-glass phase (SG) for
  $T\!<\!T_\mathrm{c}(h)$. For $T>T_\mathrm{c}(h)$ the system is in the
  paramagnetic phase (PM). The critical line separating both phases,
  $T\!=\!T_\mathrm{c}(h)$, is called the deAlmeida-Thouless (dAT)
  line~\cite{dealmeida:78}. The value of the critical field at $T\!=\!0$ is
  called $h_{\mathrm{AT}}$. (\textbf{b}) Instead, for spatial dimension $D\!<\!D_{\mathrm{\ell}}^h$, a
  spin-glass phase exists only for $h\!=\!0$.
  According to the droplet picture, figure (b) applies in all dimensions, which corresponds to $D_{\mathrm{\ell}}^h \!=\! \infty$.}
 \label{figH:dAT}
\end{figure}
One of the most striking predictions of the RSB solution of the
SK model~\cite{sherrington:75} is a line of transitions
in a magnetic field terminating in the zero field transition point, $T_\mathrm{c}$,
see \fref{figH:dAT}(a). This was first found by de Almeida and
Thouless~\cite{dealmeida:78}
and so is known as the dAT line\footnote{Frequently this has been referred to as
the AT line, but here we indicate correctly the initials of the first
author.}. Below the dAT line, the SK model is described by the RSB solution of
Parisi~\cite{parisi:79,parisi:80,parisi:83}, while above the dAT line the replica symmetric solution is
valid. According to RSB theory, there is also a dAT line in short-range models.
If there is no dAT line, then there is simply a line of transitions
along the zero field axis, terminating at $T_\mathrm{c}$, as shown in \fref{figH:dAT}(b). This is the prediction of the droplet theory. 

While the zero field transition has a spontaneously broken symmetry, as usual,
the transition in a field on the dAT line is \textit{unusual} in having no broken
symmetry, since spin-inversion symmetry is already broken by the magnetic
field. While the dAT transition definitely occurs in the SK model, there is
controversy as to whether it also occurs in models which do not have
infinite-range interactions. According to RSB theory there is a dAT line in short-range models, whereas according to the droplet theory the dAT line is an artefact of the infinite-range nature of the SK model and does not occur in short-range models in \textit{any} dimension.

In critical phenomena, we are familiar with the fact that fluctuations
destroy a transition in dimension $D$ below a ``lower critical
dimension", $D_{\mathrm{\ell}}$, where $D_{\mathrm{\ell}}\!=\!1$ for the Ising
ferromagnet (see
e.g. Ref.~\cite{parisi:88}),
and $D_{\mathrm{\ell}}\!\approx \! 2.5$ for the Ising
spin-glass~\cite{boettcher:05,franz:94,maiorano:18}. What about the spin glass
in a magnetic field?
The gauge symmetry in Eq.~\eqref{eq:Gauge-Transform} implies that
even a uniform magnetic field in a spin glass is effectively a
random-field as far as the spin glass ordering is concerned.
This observation reminds us immediately of the random field
Ising model (RFIM, see e.g.~\cite{nattermann:98}).

When random-fields are switched on, they energetically favor spin configurations
which are completely unrelated to the ferromagnetic (or spin-glass)
ordered configurations that one finds in the absence of a field.
The outcome
of this competition crucially depends on the space dimension. If
$D\!>\!D_{\mathrm{\ell}}^h$, the low-temperature ordered phase survives in the
presence of small
random-fields, while if $D\!<\!D_{\mathrm{\ell}}^h$, the
slightest random-field destabilizes the ordered phase, see \fref{figH:dAT}(b).
Clearly $D_{\mathrm{\ell}}^h\geq D_{\mathrm{\ell}}$ and indeed for the RFIM,
$D_{\mathrm{\ell}}^h\!=\!2$, which is greater than $D_{\mathrm{\ell}} \!=\! 1$. Unfortunately, a consensus has not yet emerged about the value of $D_{\mathrm{\ell}}^h$ for the spin glass. Here will discuss some numerical attempts to determine it.

de Almeida and Thouless~\cite{dealmeida:78} computed the stability of the replica symmetric
solution of the SK model, finding that an eigenvalue went negative (indicating
instability) below the dAT line. Fortunately, this instability can be located
in simulations because the unstable (``replicon")
eigenvalue is the inverse of the spin glass susceptibility
$\chi_{SG}$ in the presence of a magnetic field, recall Eq.~\eqref{eq:chi-SG}.

Hence the goal is to locate a divergence in $\chi_{SG}$. Of course no
divergence occurs in the finite systems which are simulated, so we need to
locate the dAT line by finite-size scaling (FSS), recall
Sect.~\ref{subsect:FSS}. This is most straightforward using a dimensionless
quantity such as $\xi_2/L$ where $\xi_2$ is the correlation length of a finite
system (second-moment correlation length), defined in
Eq.~\eqref{eq:xi2-def}. Now we recall Eq.~\eqref{eq:xi-FSS}, which indicates that 
the data for
$\xi_2/L$ for different sizes intersects at the transition and splays out again
on the low temperature side. We therefore look for intersections in the data.

\begin{figure}[t]
\centering
\includegraphics[width=6cm]{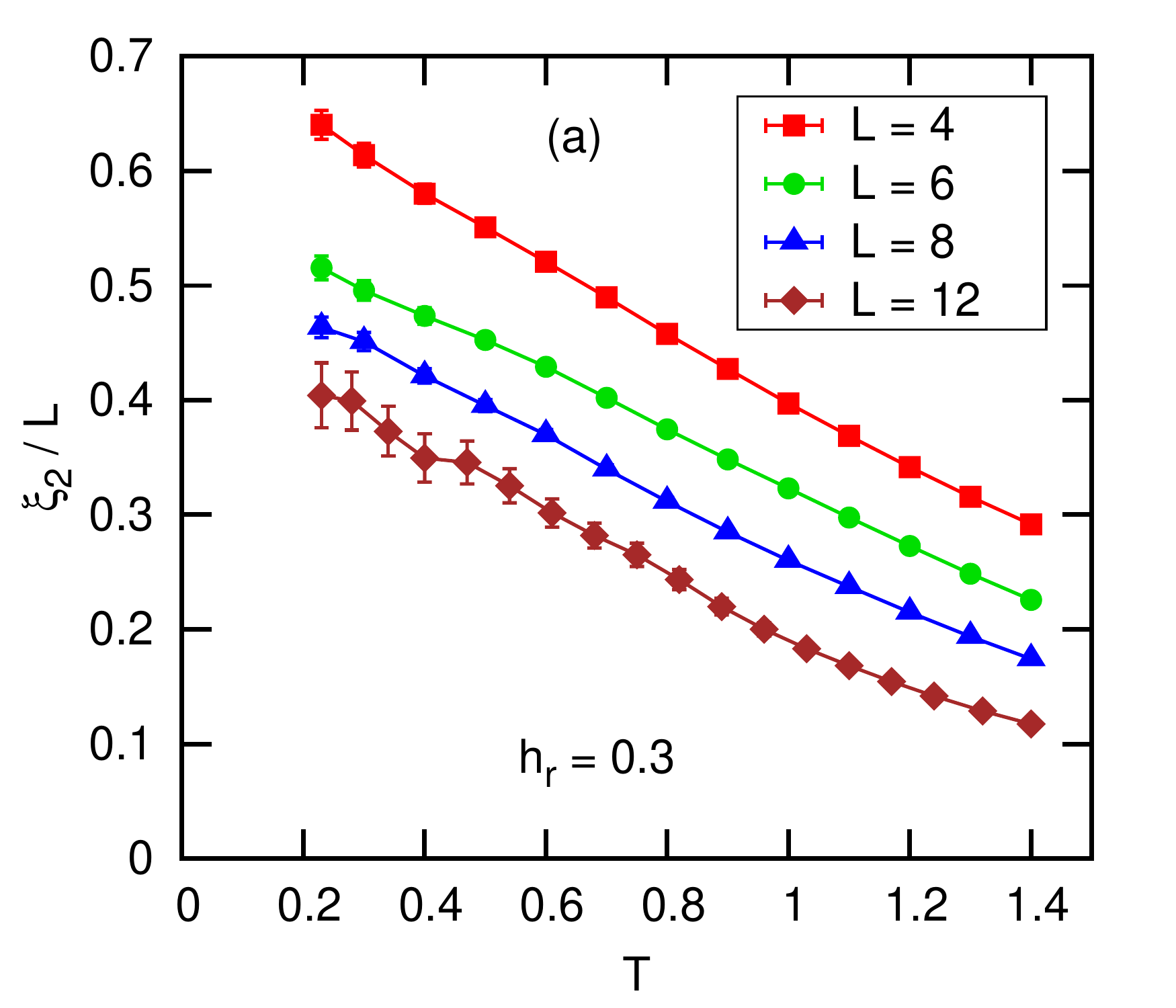}
\includegraphics[width=6cm]{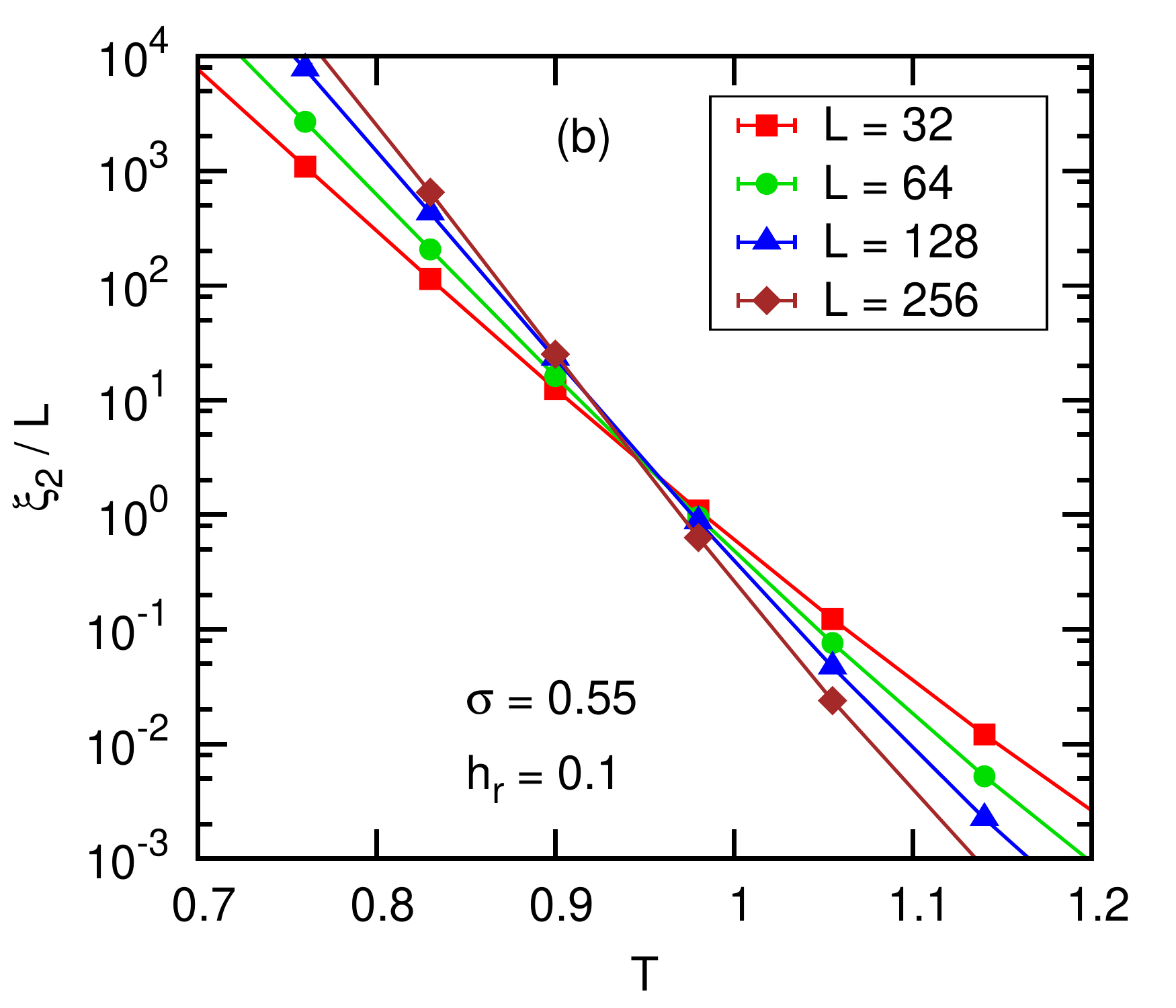}
\caption{(\textbf{a}): Temperature dependence of the second-moment correlation length in
  units of the lattice size, $\xi_2/L$, as computed in spatial dimension $D\!=\!3$
  for cubic samples of linear size $L$. The computation was carried out for
  Gaussian-distributed couplings and random fields, with zero mean and
  standard deviations $\sigma_J\!=\!1$ (this actually sets the energy units) and
  $\sigma_h\!=\!h_\mathrm{r}\!=\!0.3$.  The lack of intersections down to the lowest
  temperature $T\!=\!0.23$ suggests that $D\!=\!3$ lies below the lower-critical
  dimension in a field, recall~\fref{figH:dAT}. The figure is adapted from Ref.~\cite{young:04}.
  (\textbf{b}): Second moment correlation length for a model with long-range
  power-law interactions for parameters corresponding to a short-range model
  in dimension $D$ greater than 6, which is the ``upper critical dimension"
  for the spin glass in zero field.
  The intersections suggest that there \textit{is} a dAT line in this region.
  Adapted from Ref.~\cite{katzgraber:05}.}
 \label{figH:xiLoverLYK2004}
\end{figure}
Early studies did not find a spin glass transition in a field in $D\!=\!3$
using standard FSS methods~\cite{young:04,jorg:08b}, see \fref{figH:xiLoverLYK2004}(a). It is difficult to
directly simulate spin glasses in high dimensions, because the number of sites
$N\!=\! L^D$ increases so fast with linear size $L$, that one can not equilibrate
enough values of $L$ to perform FSS. Instead, it has been proposed to study models in 1$D$ with long-range interactions which fall off as a power $\sigma$, and, for each
value of the power, the model serves as a proxy for a short-range model in
a dimension $D$ which depends on $\sigma$. Figure~\ref{figH:xiLoverLYK2004}(b)
shows a plot from data in Ref.~\cite{katzgraber:05} for the long-range model
parameters corresponding to a short-range model for $D\!>\!6$. Intersections are
clearly seen indicating a dAT line in this region. Varying the power $\sigma$,
the data of Ref.~\cite{katzgraber:05} shows intersections for models
corresponding to $D \!>\! 6$ but not for $D\! <\! 6$.

As pointed out in Ref.~\cite{janus:12}, a
difficulty in applying FSS to spin glasses is that extrapolating the inverse of the
wave-vector dependent propagator, Eq.~\eqref{eq:G-replicon}, to ${\boldsymbol k} \!=\! 0$ gives a different value
from $1/\chi_{SG}$ (which is the inverse of the propagator evaluated directly at ${\boldsymbol k\!=\!0}$), even away from the
transition. The usual computation of the
correlation length $\xi_2$ involves the ${\boldsymbol k}\!=\! 0$ value as well as a ${\boldsymbol k} \ne 0$ value.

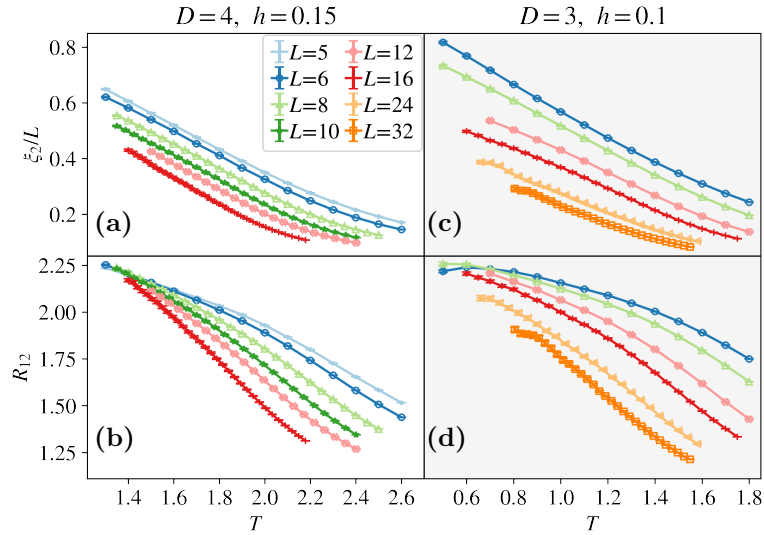
\begin{figure}[t]
\centering
\begin{overpic}[trim=0 5 0 5,clip,width=4in]{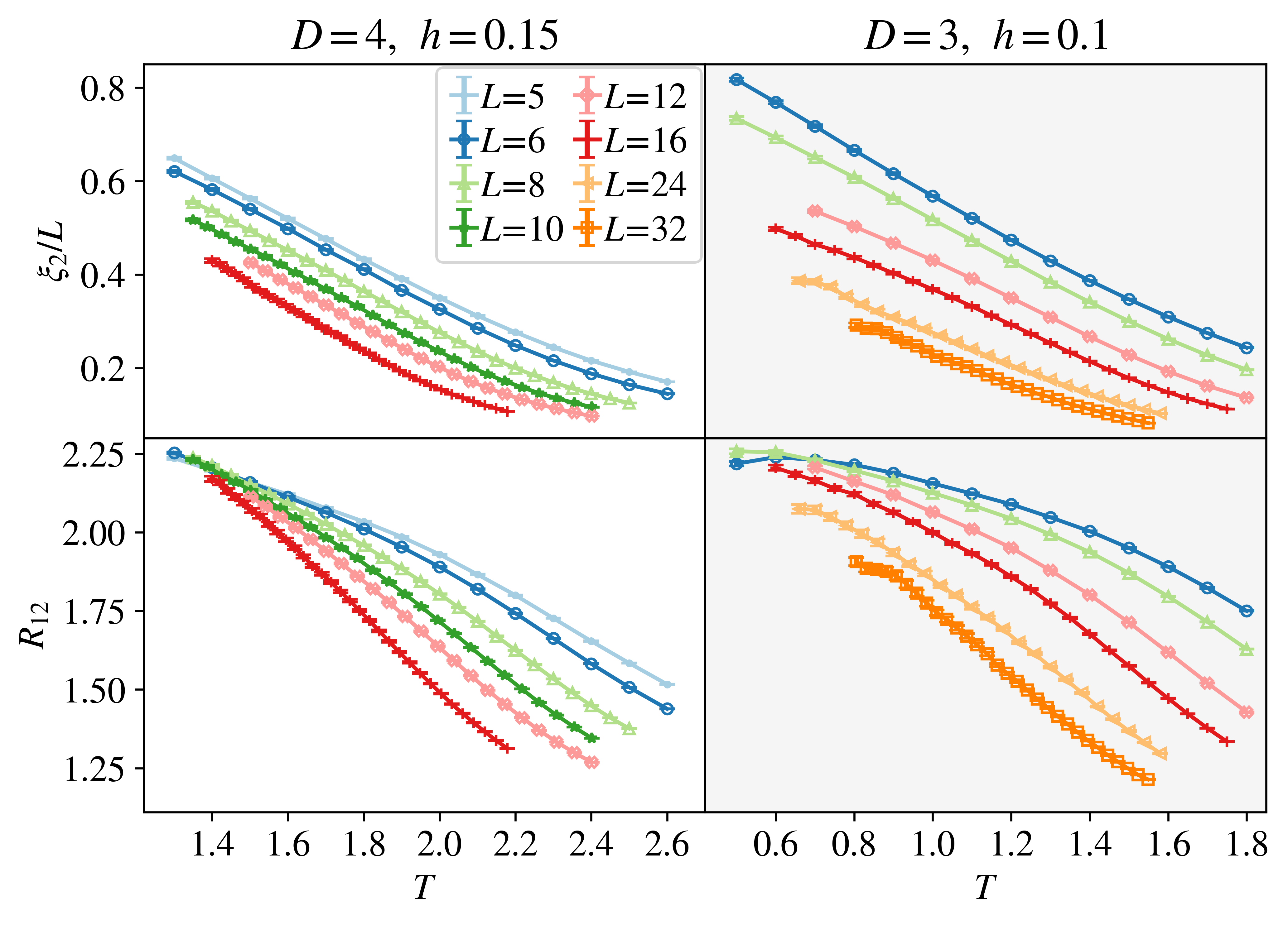}
\put(12,40){{\bf (a)}}
\put(12,12){{\bf (b)}}
\put(55,40){{\bf (c)}}
\put(55,12){{\bf (d)}}
\end{overpic}
\caption{Data for $D\!=\!4$ taken from Ref.~\cite{janus:12}, and $D\!=\!3$, from Ref.~\cite{janus:14c}. Top: a plot of $\xi_2/L$ as a function of temperature for $D\!=\!4$ and $h\!=\!0.15$ (\textbf{a}), and  $D\!=\!3$ and $h\!=\!0.1$ (\textbf{c}). According to leading-order finite-size scaling, the curves for different sizes should
intersect at the phase transition point.
conclude that there is no phase transition in this system. 
Bottom: Plot of the dimensionless ratio $R_{12}$, Eq.~\eqref{eq:R12}, for the
same parameters [(\textbf{b}): $D\!=\!4$, (\textbf{d}): $D\!=\!3$] , which should have the same leading-order scaling as $\xi_2/L$. Unlike the correlation length, however, $R_{12}$ does exhibit intersections in $D\!=\!4$, which suggests that
there is a phase transition in a field. In $D\!=\!3$ the only intersection is for the smallest sizes. It is not known if there would be intersections at larger sizes and lower temperatures than can be simulated.  
}
\label{figH:xiLoverL-R12}
\end{figure}

In $D\!=\!4$, Ref.~\cite{janus:12} used both standard FSS to compute $\xi_2$ and a
non-standard approach to compute a different dimensionless quantity $R_{12}$, defined in Eq.~\eqref{eq:R12},
which does not involve the ${\boldsymbol k} \!=\! 0$ data point. The data for $\xi_2/L$
does not show a transition while that for $R_{12}$ does, see 
\fref{figH:xiLoverL-R12}--left panel. This may indicate a dAT line in $D\!=\!4$, but it is
disappointing that two methods, which should give the same result in the
asymptotic limit, give different results for the sizes that can be simulated. It has been argued~\cite{leuzzi:09} that it is preferable to avoid ${\boldsymbol k}\!=\! 0$ because it has large corrections to FSS coming from the the negative-$q$ region of $P(q)$. On the other hand the ${\boldsymbol k}\!=\! 0$ point is the most divergent, which one would therefore \textit{normally} like to include, so at present there is not a general consensus in the community on whether or not a phase transition occurs in a field in $D\!=\!4$.

In $D\!=\!3$, even avoiding the anomaly in the propagator did not
result in any evidence for a phase transition in the presence of a field, see
Ref.~\cite{janus:14c} and \fref{figH:xiLoverL-R12}--right panel.

To conclude this section, despite a huge computational effort and careful analysis, the range of
dimensions in which there is a dAT line has not been demonstrated
convincingly. The problem is that corrections to FSS are large and not
adequately understood\footnote{We have already noted the discrepancy between 
$\lim_{{\boldsymbol k}\to 0}\hat G^{-1}_{SG}({\boldsymbol k})$ and $\chi^{-1}_{SG}$.}. In zero field,
the work of Refs.~\cite{hasenbusch:08b,janus:13} determined the exponent
for the largest correction to scaling and the amplitude of those correction terms.
One is then confident that the data is in the asymptotic scaling region.
However, in a field it has not been possible to identify, and hence compensate
for, corrections to scaling in the presence of a field. 

Several interpretations
of the numerical results on spin glasses in a field are possible:
\begin{itemize}
\item One possibility is that $3 \!<\!  D_{\mathrm{\ell}}^h \!<\! 4$. In fact,
  some RSB computations suggest~\cite{parisi:12} that $D_{\mathrm{\ell}}^h \!<\! 6$.
  Unfortunately, these analytic computations do not provide a precise estimate
  for $D_{\mathrm{\ell}}^h$.
\item Other analytic computations suggest that
  $D_{\mathrm{\ell}}^h\!=\!6$~\cite{yeo:15}. From this point of view, the observation
  of scale-invariance in some of the $D\!=\!4$ simulation results is attributed to a
  correlation length which is enormous, but
  finite, in the thermodynamic limit~\cite{aspelmeier:16}.
  A possible piece of evidence in favor of this hypothesis is that a
  renormalization group calculation of Bray and
  Roberts\cite{bray:80b} found a fixed point for the dAT line in $D > 6$ but not
  in $D \!<\! 6$.\footnote{This means that the upper critical dimensions of the model is also $D^h_u\!=\!6$. However a recent analytical computation claims that $D^h_u\!=\!8$.~\cite{angelini:22}} 
\item A more exotic possibility is that $D_{\mathrm{\ell}}^h\approx 3$, but
  the phase-transition in a field is hidden by truly dramatic
  statistical fluctuations~\cite{parisi:12b,janus:14c}.
  
\item Another possibility is that of a quasi-first order
  transition~\cite{holler:20} (which, thinking in retrospect, would explain
  naturally many of the dynamic findings in a $D\!=\!3$
  simulation~\cite{janus:14b}). Yet, equilibrium data in the $D\!=\!4$ scaling
  region do not seem to conform to this expectation~\cite{fernandez:22}.

\end{itemize}
We hope that time will tell us which of the above possibilities (if any!) is
an accurate description of the spin-glass phase diagram. Since a huge
computer effort has \textit{already} been expended~\cite{janus:12,janus:14c},
significant future progress is likely to need new ideas as well as more
computer power.

\section{Out-of-equilibrium}\label{sect:out-equilibrium}
Numerical studies of out-of-equilibrium behavior are designed to mimic experiments, 
and have the advantage over experiments that they can track the microscopic evolution of the system. In this section, 
we shall consider the simplest aging experiment, in which
a very large system is instantly cooled from $T=\infty$ to $T<\Tc$,\footnote{$L$ should be much larger than the growing spin glass coherence length $\xi(t)$, in order to avoid finite-size effects.} and its microscopic evolution is followed as function of $\tw$, the waiting time elapsed after the quench. In some cases, in order to
reproduce an experimental protocol, a small magnetic field $h>0$ will be switched on at time $\tw$, and the system's response to the field measured at a later time $t+\tw$. In these simulations and experiments, the magnetic field is viewed \emph{only} as probe of the $h=0$ spin-glass state. 

In this context, and at variance with equilibrium studies,
one is interested in the evolution of an $L\to\infty$ system, at finite times $\tw$ and $t$, so the limit $L\to\infty$ must be taken \emph{before} $\tw$ and $t$ get large. Recent simulations explore
a time range going from the equivalent of picoseconds to tenths of a second, while the time range in experiments goes from seconds to of order 24 hours. Unfortunately, neat theoretical predictions apply only in the limit of very long $\tw$. We emphasize that the real controlling variable is not time, but the size of the glassy domains, which we quantify through the spin-glass coherence length $\xi(t)$, see Eq.~\eqref{eq:replicon} below.

\subsection{Observables (out-of-equilibrium)}\label{subsect:obs-off-equilibrium}
\begin{figure}[h]
\centering
\begin{overpic}[height=1.7in]{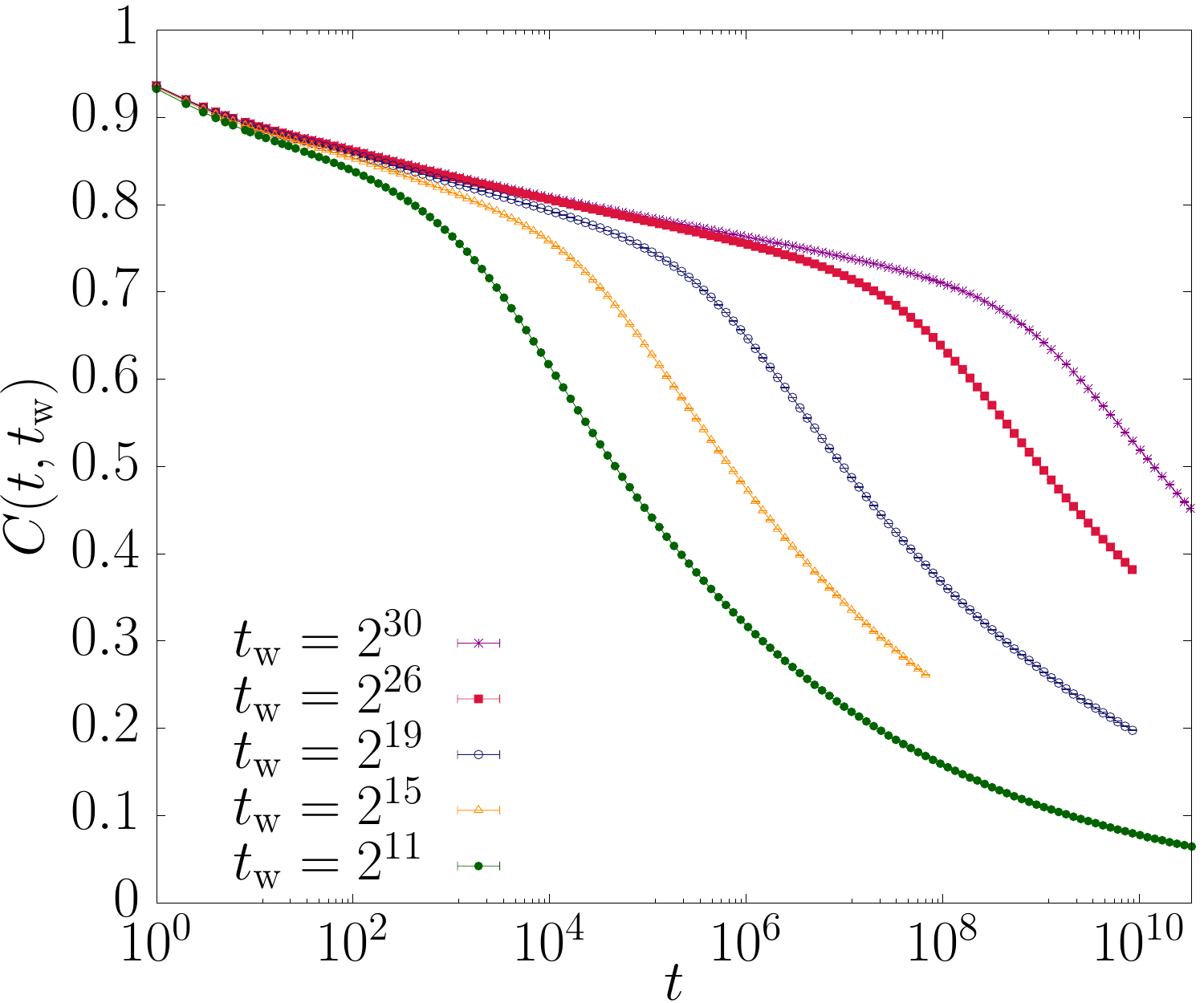}
\put(50,75){{\bf (a)}}
\end{overpic}
\begin{overpic}[height=1.7in]{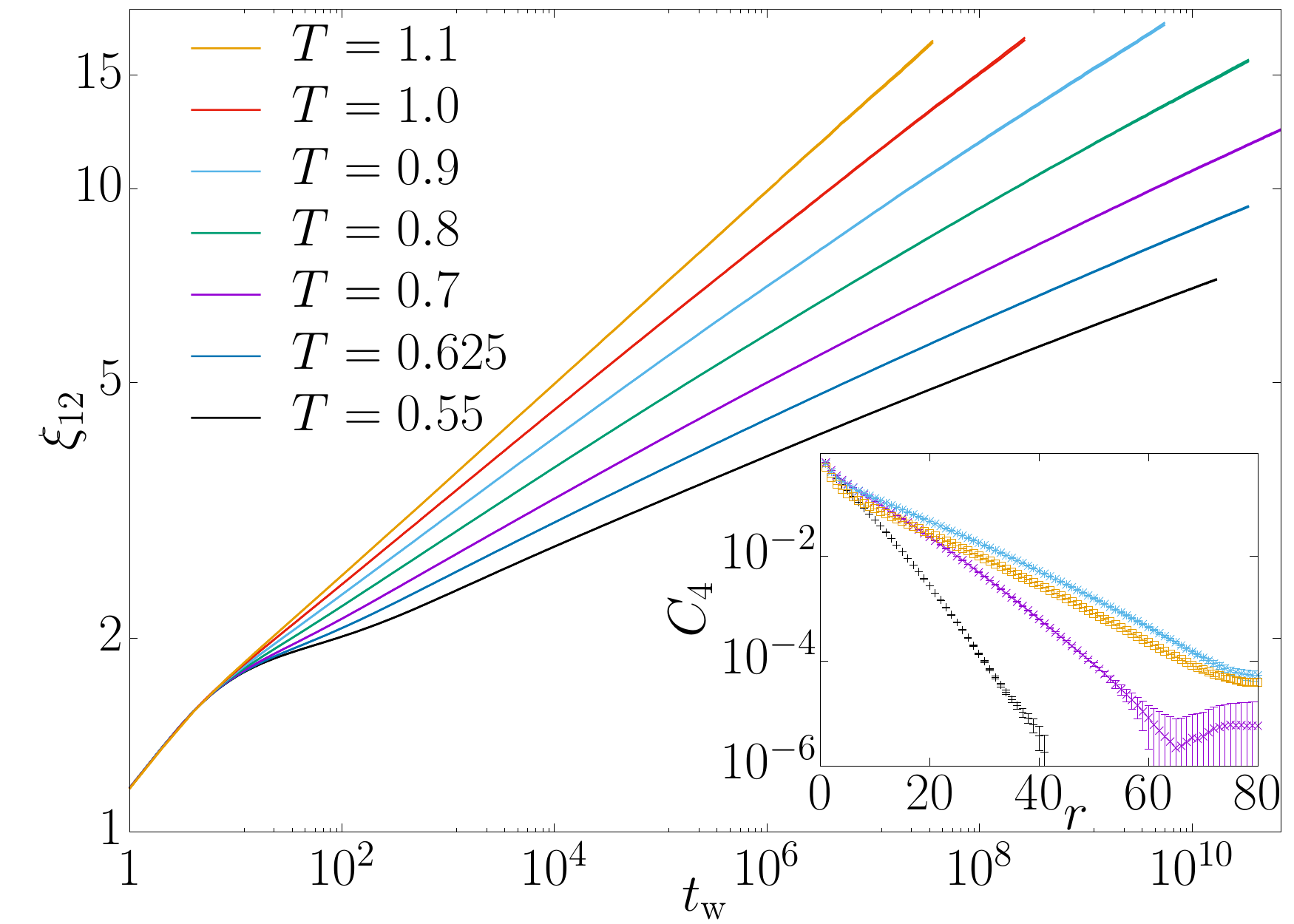}
\put(45,65){{\bf (b)}}
\put(85,30){{\bf (c)}}
\end{overpic}
\caption{Aging dynamics for the $3D$ EA model with couplings $J=\pm 1$. (\textbf{a}) Decay of the two-time spin correlation function $C(t+\tw,\tw)$, see Eq.~\eqref{eq:C}, for five different values of $\tw$ ranging from 2048 up to $10^8$ Monte Carlo sweeps, at $T=0.7$ (data taken from~\cite{janus:17}). (\textbf{b}) Size of the glassy domains, as quantified by the spin-glass coherence length $\xi(\tw)$, as function of $\tw$, for different temperatures $T$. For comparison, the critical temperature for this model is $T_\mathrm{c}=1.1019(29)$~\cite{janus:13}.  (\textbf{c}): Autocorrelation function $C_4(r,\tw)$, see Eq.~\eqref{eq:C4}, as a function of distance $r$, as computed for the longest $\tw$  for each temperature ($T$ color key as in main panel). Note that $C_4$ varies by six orders of magnitude in this computation.  Figure taken from~\cite{baity-jesi2018}.}\label{fig:xitw}
\end{figure}

An important and striking observation is that the older a spin glass is (i.e.~the longer it has waited), the slower its subsequent relaxation becomes. This is called \emph{aging}. Aging can be studied with the two-time spin correlation function, see Fig.~\ref{fig:xitw}(a),
\begin{equation}\label{eq:C}
C(t+t_\mathrm{w},t_\mathrm{w})=\left[\left\langle S(\boldsymbol{x},t+t_\mathrm{w}) S(\boldsymbol{x},t_\mathrm{w})\right\rangle_T\right]\,,
\end{equation}
where the thermal noise average $\left\langle\cdots\right\rangle_T$
represents an average over independent thermal histories at temperature $T$.

Aging dynamics is directly related to the sluggish growth of glassy order with $\tw$, see Fig.~\ref{fig:xitw}(b). The size of spin glass domains at $\tw$, namely the coherence length, $\xi(\tw)$, is extracted with high precision (see Fig.~\ref{fig:xitw}) in simulations from the decay of the spatial autocorrelation function of the overlap field,
\begin{align}\label{eq:C4}
C_4(\boldsymbol r, \tw) &= \left[\left\langle q^{(a,b)}(\boldsymbol x, \tw)
q^{(a,b)}(\boldsymbol x+\boldsymbol r , \tw)\right\rangle_T\right],
\end{align}
where
\begin{align}
q^{(a,b)}(\boldsymbol x,\tw) &= S^{(a)}(\boldsymbol x,\tw)S^{(b)}(\boldsymbol x,\tw)\,.
\end{align}
$C_4(\boldsymbol r, \tw)$ displays scaling behavior at long distances,
\begin{equation}\label{eq:replicon}
 C_4(r,\tw) = r^{-\theta} f\bigl(r/\xi(\tw)\bigr),
\end{equation}
from which one can determine $C_4$ and the ``replicon" exponent $\theta$. The problem of extracting $\xi(\tw)$ without knowing the precise form of the scaling function $f$ has been circumvented using integral estimators (see e.g.~\cite{belletti2009depth} and the supplemental material for Ref.~\cite{baity-jesi2018}).  However, $\xi(\tw)$ is only directly accessible in simulations. In order to
estimate it from experiments we need to take an indirect route by perturbing the system with a magnetic field.

In the \emph{zero-field cooled} protocol, the only
one considered here, the field is switched-on at time $\tw$ and the magnetization density $m(t+\tw)=\sum_{\boldsymbol x} S_{\boldsymbol x}(t+\tw)/N$ is studied
as function of $h$, $t$ and $\tw$ with the initial condition $m(\tw)=0$. We have
\begin{align}\label{eq:susc}
m(t+t_\mathrm{w})&=\chi(t+\tw,t_\mathrm{w}) h\ -\ \chi_3 (t+\tw,t_\mathrm{w}) \frac{h^3}{3!}+\ldots\, \\
S(t,\tw;h)&=\frac{1}{h}
\frac{\partial m(t+t_\mathrm{w})}{\partial\log t}\,.
\label{eq:SttwH}
\end{align}
which defines the linear ($\chi$), and non-linear ($\chi_3,\chi_5$, etc.) susceptibilities, as well as the response function $S(t,\tw;h)$. In equilibrium, the fluctuation dissipation theorem (FDT) relates $\chi$ to the two-time correlation function $C(t+\tw,t_\mathrm{w})$ by $T\chi=1-C$.
Out-of-equilibrium, the relationship between $\chi$ and $C$ is even more interesting (see Sec.~\ref{subsect:FDR}).

The size of the glassy domains is experimentally accessed through the relaxation function $S(t,\tw;h)$, see Eq.~\eqref{eq:SttwH}, that peaks at an effective time $t^\mathrm{eff}(h)$. The effective time gets shorter when
the magnetic field increases, due to the Zeeman-effect lowering the free-energy barriers. The Zeeman effect gets stronger as $\xi(t)$ grows, which can be used to experimentally measure $\xi(\tw)$, see~\cite{joh1999extraction} for details. Interestingly, this experimental set-up for determining $\xi(\tw)$ has been reproduced in simulations and found to yield values in agreement
with the microscopic determination of $\xi(\tw)$ from Eq.~\eqref{eq:replicon}~\cite{baity2017matching}.  In fact, numerical and experimental data for $t^\mathrm{eff}(h)$ can be described with the same scaling functions, see Ref.~\cite{zhai:2020} and Fig.~\ref{fig:replicon}(b).

According to simulations and experiments, $\xi(\tw)$ varies roughly as a small power of $\tw$, i.e.~ $\xi(\tw)\propto\tw^{1/z(T)}$, with an exponent that depends on the temperature $z(T)\simeq z(T_\text{c}) \Tc/T$ (see Fig.~\ref{fig:xitw} and Refs.~\cite{rieger:95, marinari:96,joh1999extraction,marinari:00b,berthier2002geometrical,belletti2009depth,fernandez2015testing,nakamae2012dynamic}). However, this behaviour is not exact and, for the range of $\xi(\tw)$ that can be studied, the exponent $z$ is found to depend slightly on
$\tw$~\cite{zhai:2017,baity-jesi2018,zhai2019slowing} as well as on $T$.
As we will discuss in Secs.~\ref{subsect:FDR} and~\ref{subsect:dictionary-statics-dynamics}, this slowly growing $\xi(\tw)$ allows us to build a quantitative \emph{statics-dynamics dictionary}, relating non-equilibrium dynamics of infinite systems at finite time $\tw$ with equilibrium properties of finite systems of size $L\sim \xi(\tw)$. Hence experiments, which are inevitably out of equilibrium, can described by the equilibrium physics of systems of size of order $\xi(\tw)$, which is not much larger than those explored in recent simulations.

\subsection{The replicon exponent}
The droplet and RSB theories disagree about the exponent in the algebraic prefactor of Eq.~\eqref{eq:replicon}, $\theta$, which is called the \emph{replicon exponent}.\footnote{It is unfortunate that the stiffness exponent, $\theta_\mathrm{S}$, is sometimes also called
$\theta$, which may cause confusion.}
 Droplet theory expects coarsening behavior, so $\theta=0$, while RSB expects $\theta>0$. Early simulations found $\theta=0.50(2)$~\cite{marinari:96}. More recent and accurate numerical simulations find $\theta$ in the range $0.35-0.4$\cite{belletti2009depth,baity-jesi2018}. From the perspective of droplet theory, see e.g.~\cite{moore2021droplet}, it has been argued that the non-vanishing value of $\theta$ is due to a transient effect in which the system feels the effects of the renormalization group fixed-point at $T_\mathrm{c}$, rather than the asymptotic fixed-point at $T=0$.

\begin{figure}[b]
\centering
\begin{overpic}[height=1.8in]{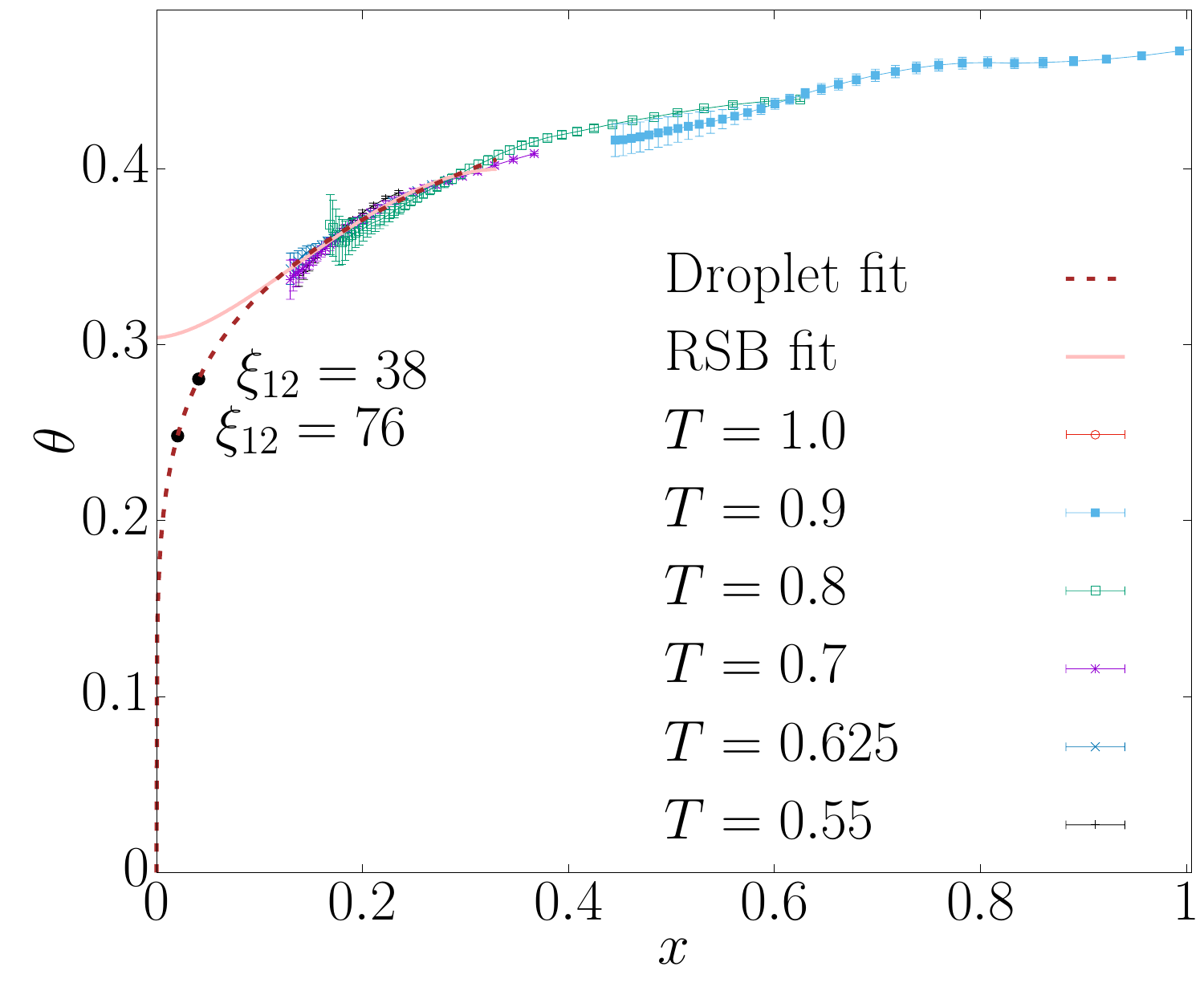}
\put(15,77){{\bf (a)}}
\end{overpic}
\begin{overpic}[height=1.8in]{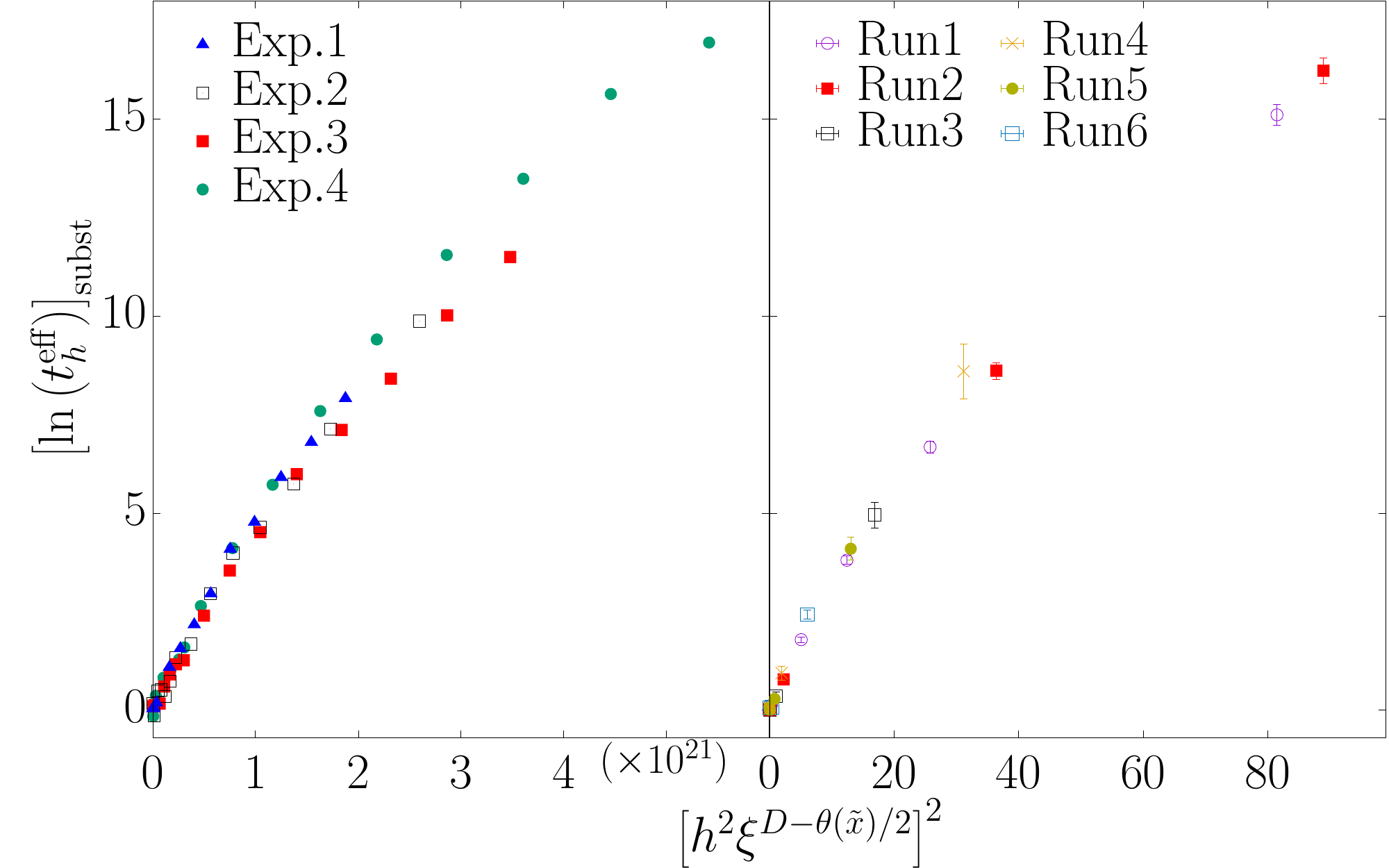}
\put(35,58){{\bf (b)}}
\end{overpic}
\caption{(\textbf{a}) When represented as a function of $x=\ell_\text{J}(T)/\xi(\tw,T)$, the replicon exponent $\theta$ obtained for several temperatures fall on a single curve. In addition we show a droplet extrapolation [dashed line, $\theta(x\to 0)=0$] and an RSB extrapolation [full line, $\theta(x\to 0)>0$]. Even the droplet extrapolation, which has $\theta(x) \to 0$ as $x \to 0$, predicts sizeable values of $\theta$ for the values of $x$ relevant to experiments (black dots). Figure adapted from~\cite{baity-jesi2018}.  (\textbf{b})  After a proper subtraction~\cite{zhai:2020}, the logarithm of the effective time [the time at which the response function~\eqref{eq:SttwH} peaks] obeys the same scaling for CuMn (left) and the Ising-Ewdards-Anderson model (right). Figure taken from~\cite{zhai:2020}.
}\label{fig:replicon}
\end{figure}

In fact, the crossover between the $T_\mathrm{c}$ and $T=0$ fixed points can be studied systematically through the ratio of the Josephson length, $\ell_\text{J}(T)\propto(T_\text{c}-T)^{-\nu}$, to $\xi(\tw)$. The asymptotic value of $\theta$ is obtained
when the ratio $x=\ell_\text{J}(T)/\xi(\tw)$ goes to zero. In fact, as shown in Fig.~\ref{fig:replicon}(a), the ratio $x$ seems to be
the controlling variable for $\theta$, which shows a
decreasing trend when $x$ goes to zero. The data are
compatible with both a vanishing (droplet) and
non-vanishing (RSB) extrapolation to  $x=0$. 
The crucial point however, is that neither simulations nor experiments are carried out at $\xi=\infty$ (i.e. $x=0$). Rather,
representative values for extrapolations to the experimental
scale are shown in Fig.~\ref{fig:replicon}(a): even the droplet scaling predicts $\theta>0.25$ at those $x$. In fact, several successful extrapolations of simulation results
to the experimental scale have been carried-out recently
\cite{baity-jesi2018,zhai2019slowing,zhai:2020} [see also Figs.~\ref{fig:replicon}(b) and~\ref{fig:simulationsmeetexperiments}], all of them with
$\theta>0$. Hence, current simulations \emph{and} experiments are all carried out in an RSB-like regime.

\subsection{Generalization of the Fluctuation-Dissipation theorem}\label{subsect:FDR}

The generalization of the FDT theorem to the out-of-equilibrium regime for the SK model~\cite{cugliandolo1993analytical} opened the possibility of checking some of the predictions of RSB. This generalization of FDT was numerically tested in the $3D$ EA model~\cite{franz1995fluctuation,marinari1998violation} and was finally proved for finite dimensional systems assuming stochastic stability~\cite{franz:98} (see Sec. \ref{subsect:replica-equiv}). 
 The generalized fluctuation-dissipation relation (GFDR) reads\footnote{Assuming a dependence  of the magnetic field with time as $h(t)=h_0 \theta(t-t_w)$.}
\begin{equation}\label{eq:GFDR}
T\chi(t+\tw,t_\mathrm{w})= S(C(t+\tw,t_\mathrm{w}))\,,
\end{equation}
where $S(C)$ is a function which can be shown~\cite{cugliandolo1993analytical,franz1995fluctuation,marinari1998violation,franz:98},  to be related, in the limit of $\tw,t\to \infty$, to a double integral of $P(q)$, 
the equilibrium pdf of the overlap, see Sec.~\ref{seq:Pq}.\footnote{This surprising result can be understood using stochastic stability. In equilibrium the system explores the region of lower free energies whereas in the out-of-equilibrium regime it wanders in the high region of the free energy. But stochastic stability tells us that the structure (maxima and minima)  of the higher and lower regions of free energy is similar.} In equilibrium, $S(C) = 1-C$, which is the FDT.

\begin{figure}[t]
\centerline{\includegraphics[width=5in,trim=30 180 30 200 ,clip]{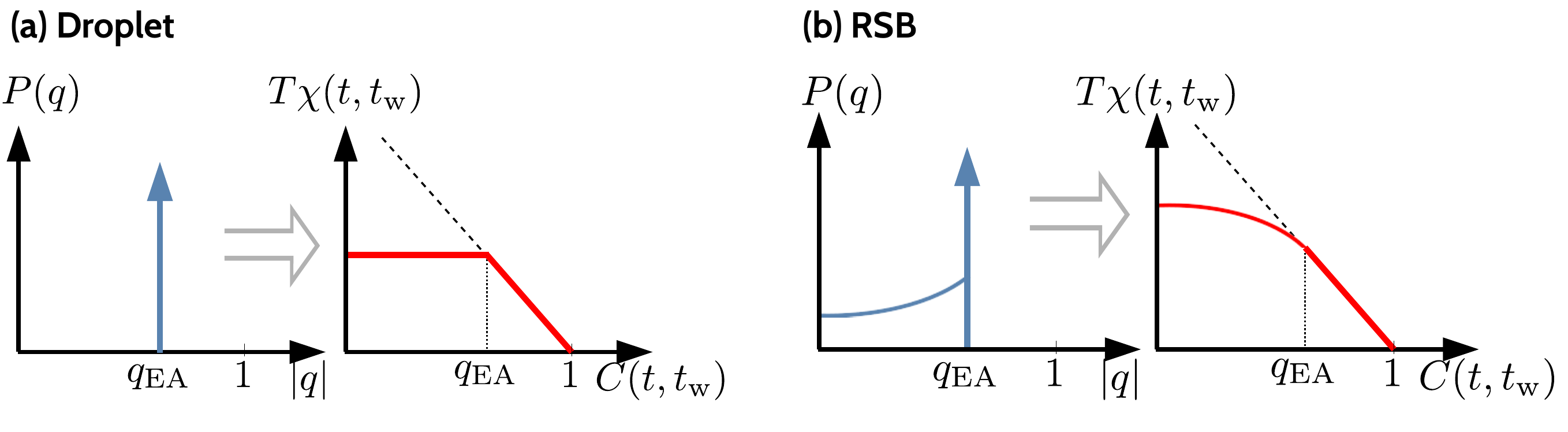}}  
\caption{Sketch of the shape of the expected modification of the FDT according to the droplet and RSB theories in the double limit of $\tw,t\to \infty$.}\label{fig:sketchFDT}
\end{figure}

\begin{figure}[t]
\begin{center}
\begin{overpic}[width=0.48\columnwidth]{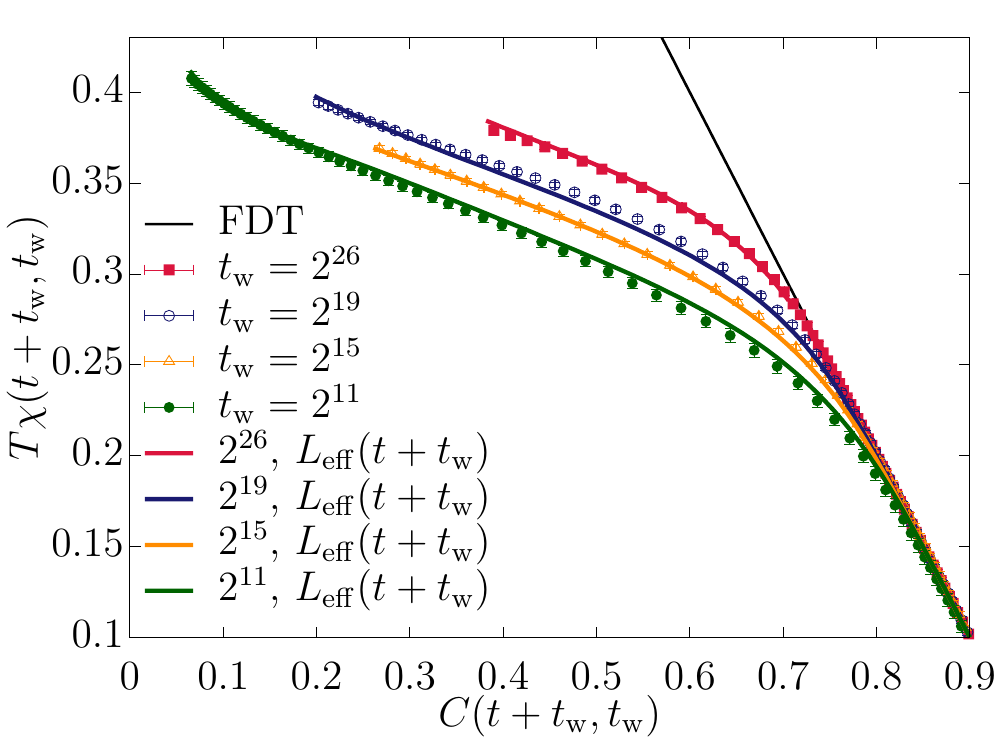}
\put(55,65){{\bf (a)}}
\end{overpic}
\begin{overpic}[width=0.48\columnwidth]{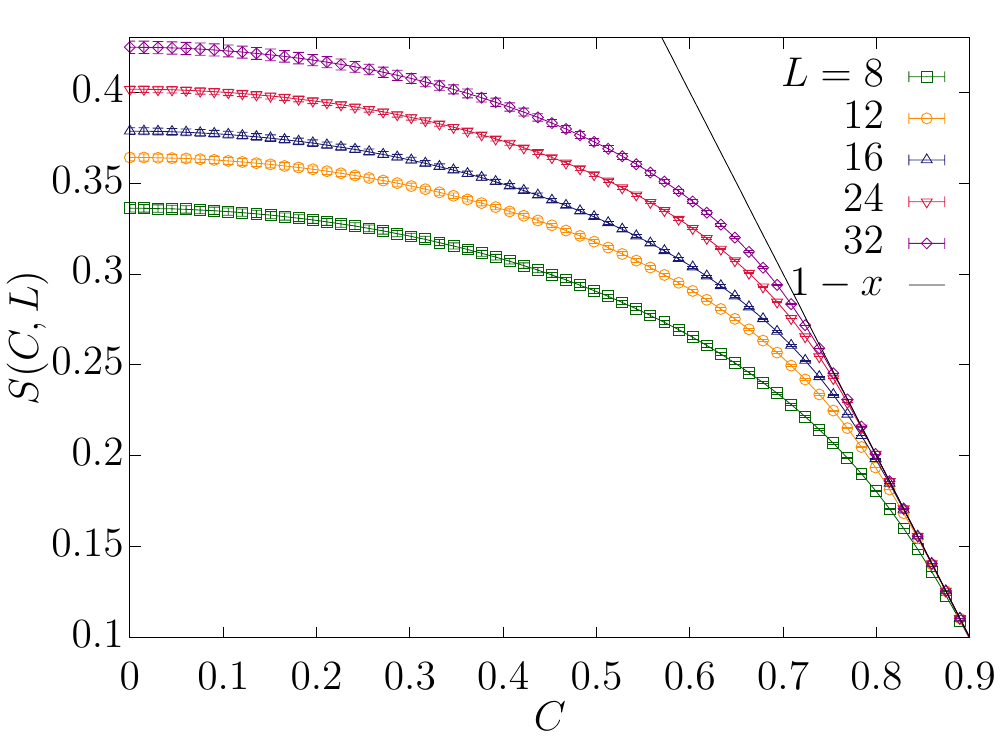}
\put(55,65){{\bf (b)}}
\end{overpic}
\end{center}
\caption{(\textbf{a}) The dots show the response function $T \chi(t+\tw,t_\mathrm{w})$ versus $C(t+\tw,t_\mathrm{w})$ at $T=0.7$ for different fixed $\tw$. Upon relaxing at fixed $\tw$, 
  $C(t+\tw,t_\mathrm{w})$ monotonically decreases from $C=1$ at $t=0$ to
  $C=0$ at $t=\infty$. In (\textbf{b}) we show the \textit{equilibrium} $S(C,L)$ versus $C$ for different system sizes obtained using Eq.~\eqref{eq:SCL} and $P(q,L)$ extracted in independent equilibrium studies. This lines are guides to the eye. In (a) the lines are obtained by plotting the $S(C,L_\mathrm{eff})$ shown in (b) for an effective size, $L_\mathrm{eff}=\xi(t+\tw,\tw) g\left(\xi(t+\tw,\tw)/\xi(\tw)\right)$, where $g(\cdot)$ is an Ansatz function that controls the crossover between the $\xi(\tw)$ and  $\xi(t+\tw)$ dominated regimes.
 Figures adapted from \cite{janus:17}.
}\label{fig:FDT}
\end{figure}

From Eq.~\eqref{eq:GFDR}, we see that the droplet and RSB theories give very different predictions about how the FDT should be modified, as sketched in Fig.\ref{fig:sketchFDT}. 
This important result allows one to the determine the pdf of the overlap experimentally, despite the impossibility of measuring it directly. Such an experiment was conducted by Hérisson and Ocio~\cite{herisson2002fluctuation} by measuring the nonlinear fluctuation-dissipation relation between $C$ and $\chi$. Many numerical works have explored these relations in simulations of Ising spin glass models~\cite{franz:98,marinari1998violation,cruz:2003,ricci2003measuring,janus:17} and beyond~\cite{crisanti2003violation}. All these experimental and numerical works describe a modification of the equilibrium FDT similar to that shown in Fig.~\ref{fig:FDT}(a) (obtained with the Janus I and II supercomputers\cite{janus:17}), which is extremely similar to the RSB prediction sketched in Fig.~\ref{fig:sketchFDT}(b). 

However it is important to recall that the connection between the equilibrium $P(q)$, and $S[C(t+\tw,t_\mathrm{w})]$ only holds in the limit $\tw,t\to \infty$. It is also clear from Fig.~\ref{fig:FDT}(a) that the current data is in a pre-asymptotic regime, because the data for different values of $\tw$ do not superimpose,
so one can not distinguish unambiguously between the droplet and RSB theories in the $\tw\to \infty$ limit. The same applies to the experimental data. This means that, as discussed above, short $\tw$ aging simulations and experiments are consistent with RSB, but cannot rule out the possibility that the droplet model describes the experimentally unachievable $\tw\to\infty$ limit.

However, it has been observed that the expected relation between $S$ and the equilibrium $P(q)$ still holds at finite $\tw$, replacing $P(q)$ by $P(q,L)$, the equilibrium pdf of the overlap in a finite system of size $L$~\cite{janus:17}. 
Indeed, Fig.~\ref{fig:FDT}(b) shows
\begin{equation}\label{eq:SCL}
S(C,L)=\int_C^1 \mathrm{d}\,C'\, x(C',L)\,,\qquad \mathrm{where}\ \ x(C,L)=\int_{0}^C\,\mathrm{d} q\, 2 P(q,L)
\end{equation}
obtained using the numerical equilibrium data of $P(q,L)$ (already discussed in Fig.~\ref{fig:Pq}), for different $L$. The similarity between the two panels of Fig.~\ref{fig:FDT} is striking since the left panel is for a non-equilibrium situation on a large system, while the right panel is for equilibrium behavior on fairly small systems. 

This discussion can be made more quantitative if we now bring into play $\xi(\tw)$, the size of the glassy domains in the non-equilibrium system that grows with time as the system ages. 
One can use the known growth of $\xi(\tw)$ (shown in Fig.~\ref{fig:xitw}), to predict the observed data from the equilibrium $S(C,L_\mathrm{eff})$ with $L_\mathrm{eff}(t+\tw;\tw)$ being a function of $\xi(\tw)$ and $\xi(t+\tw)$. This prediction is shown by the lines in Fig.~\ref{fig:FDT}(a) and the agreement with the original data (dots) is very good.

\subsection{Statics-dynamics dictionary and relation to experiments}\label{subsect:dictionary-statics-dynamics}

We have argued above that there is a quantitative equivalence between the relaxation and response of an infinite system at finite $\tw$ [and  coherence length $\xi(\tw)$] and the equilibrium properties of a system of finite size $L$. Moreover, it is possible to show that $L \sim k \xi(\tw)$, with  $k\sim 4$~\cite{janus:17}\footnote{The actual proportionality factor $k$ will depend on the details of the definition of $\xi$~\cite{belletti2009depth}. If $t$ considerably exceeds $\tw$ then $\xi(\tw)$ should be replaced by $\xi(t+\tw)$}. 

The existence of such a dictionary between equilibrium and non-equilibrium physics had also been explored and confirmed in previous works~\cite{barrat2001real,janus:08b,janus:10,janus:10b}, and 
tells us that the key to describing real experiments may not be in the physics of $L\to\infty$ equilibrium systems, but in that of much more modest sizes. 

\begin{figure}[t]
\centerline{\includegraphics[width=3in]{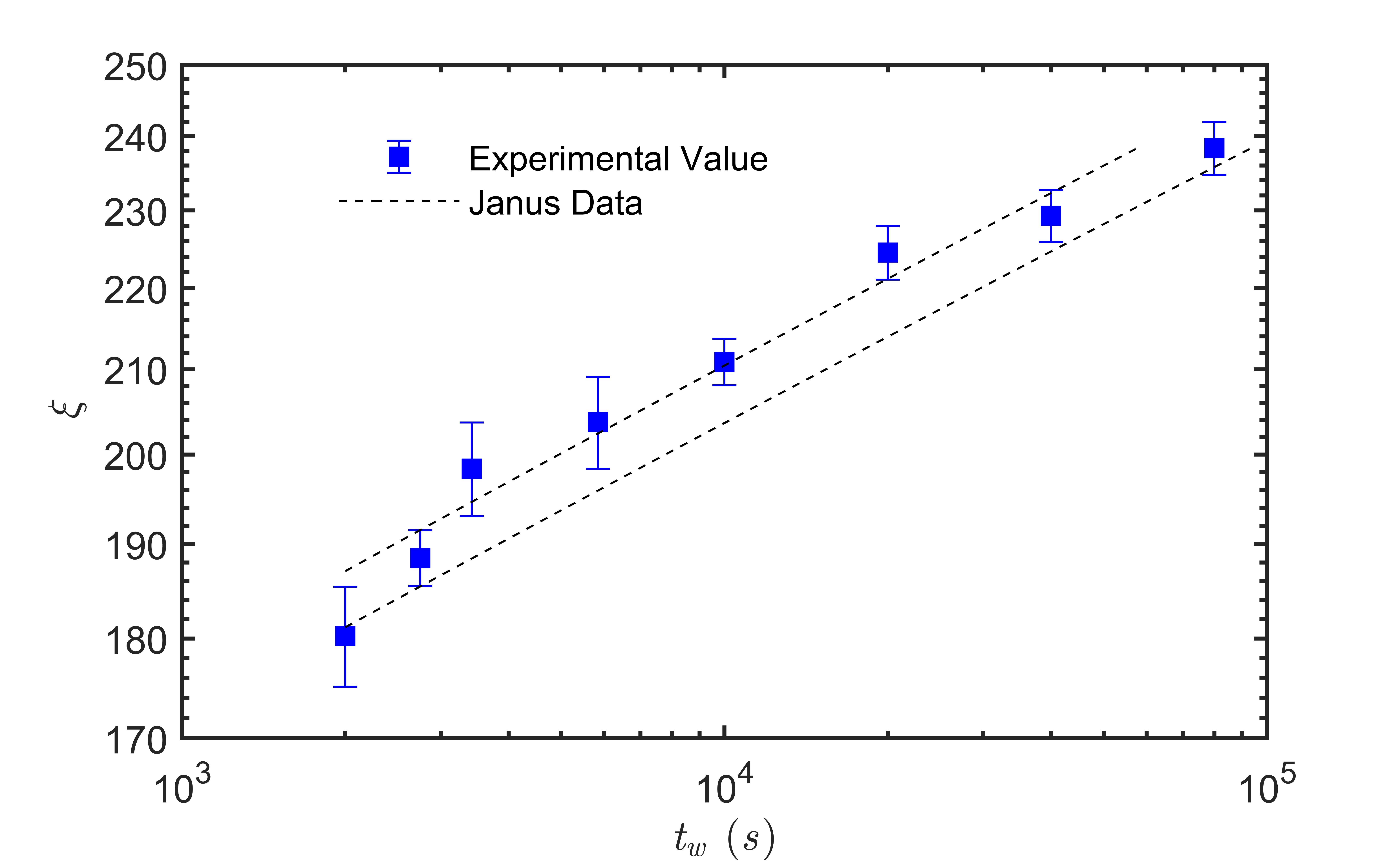}} 
\caption{Coherence length $\xi(\tw)$ in units of the average distance between magnetic moments ($0.64\, \mathrm{nm}$) versus waiting time $\tw$, as measured in a CuMn single crystal at $T\approx 0.89 T_\mathrm{c}$. The extrapolation from Janus II data at smaller $\xi(\tw)$ is enclosed by the two dashed lines (the separation between the lines is the uncertainty in the extrapolation). Figure taken from Ref.~\cite{zhai2019slowing}.
}\label{fig:simulationsmeetexperiments}
\end{figure}

The question that naturally comes to mind is how much
$\xi(\tw)$ differs in simulations and experiments. The answer is not very much. The current experimental world record for the largest $\xi(\tw)$, see
Fig.~\ref{fig:simulationsmeetexperiments}, is larger than the numerical one in Fig.~\ref{fig:xitw}(b) by a factor\footnote{Part of the factor of 15 comes from the larger values of $\tw$ in experiment, up to order $10^5$ seconds as opposed to tenths of a second in simulations, and part comes from the amplitude of the growth of $\xi(\tw)$ being larger in the experiments on CuMn than in the simulations.} $\simeq 15$.
Thus the extrapolations needed to compare experiments with current simulations, if expressed in terms of $\xi(\tw)$ rather than time, are quite mild.

In fact there is an even better quantity than $\xi(\tw)$ to study when extrapolating from simulations to experiment.
Roughly speaking we have\footnote{It is convenient to incorporate the factor of $T/T_c$.}  $\xi(\tw) = A \tw^{T/(T_\mathrm{c} z_\mathrm{c})}$. However, the data does not actually fit a power law well so it is more convenient to consider the slope on a log-log plot, which is called the aging rate, $z_\mathrm{c}(T,\xi)=(T/T_\mathrm{c})\mathrm{d}\log \tw/\mathrm{d}\log\xi$. It is found that $z_c$ is not constant but increases as $\xi(\tw)$ increases. Nonetheless, it is the most useful quantity with which to extrapolate between simulations and experimental results.

One example of such extrapolation is shown in Fig.~\ref{fig:simulationsmeetexperiments} which is taken from~\cite{zhai2019slowing} (other successful extrapolations can be found in~\cite{baity-jesi2018,zhai:2020,zhai-janus:21}). Fitting the data in the figure to a straight line gives~\cite{zhai2019slowing} $z_\mathrm{c}=12.37\pm 1.07$. Simulation results for $z_c(T, \xi)$ have been obtained for a range of waiting times and temperatures~\cite{baity-jesi2018}. For example, at the smallest correlation length, $z_c \simeq 6.7$. By fitting these simulation results as described in Ref.~\cite{zhai2019slowing} one can extrapolate the values for $z_c$ to the larger coherence lengths in experiment, getting $z_c(180.26 a_0) = 11.94\pm 0.08 , z_c(238.34 a_0) = 12.76 \pm 0.08$, which is in excellent agreement with value from the experimental data itself. Furthermore, combining the extrapolated
aging-rate with an additional input from experiment, namely $\xi(\tw=2750\, \text{s})$, the curve $\xi(\tw)$ could be predicted [the two dashed lines in \fref{fig:simulationsmeetexperiments} encompass the experimental error for $\xi(\tw=2750\, \text{s})$].

In summary, the statics-dynamics dictionary tells us that the dynamics of experimental spin glasses is described by the equilibrium properties of systems of size $L\sim 100$ lattice spacings. Hence the droplet-RSB dispute is irrelevant in this context since it applies to $L=\infty$. At the length scales that \emph{are} relevant to experiments the data is better described by RSB theory.

\section{Conclusions}\label{sect:conclusions}

The aim of this chapter has been to summarize what numerical
simulations tell us concerning the applicability of RSB to short-range
spin glasses, most particularly to $D=3$. We have seen that the
situation is somewhat complicated because it appears to depend on
whether or not an external magnetic field is applied.

In zero field, large-scale simulations with a dedicated processor find
behavior in $D=3$ which is well described by RSB, both in equilibrium
and non-equilibrium situations. The non-equilibrium simulations are
found to be consistent with recent experiments on single crystals.

While the droplet picture can not be excluded as a description of spin
glasses in the thermodynamic limit, if this is the case it can only
apply on length and time scales far beyond the reach of simulations
and experiments.

In a field, RSB predicts a spin glass phase and a dAT line, see
Fig.~\ref{figH:dAT}, whereas the droplet picture predicts no dAT
line. Numerically it has been hard to find evidence for a dAT line in
$D=3$, and the situation is ambiguous in $D=4$. Numerics does,
however, indicate the presence of a dAT line in high dimension. A
possible explanation of these results is that $D_{\ell}^h$, the lower
critical dimension, the dimension below which there is no transition,
is greater than $3$ in the presence of a magnetic field, which is
different from the case of zero field where $D_{\ell} \simeq 2.5$.

\section{Acknowledgments}

This work was partly supported by grants No.~PID2020-112936GB-I00 and
No.~PGC2018-094684-B-C21 funded by Ministerio de Econom\'{\i}a y
Competitividad, Agencia Estatal de Investigaci\'on and Fondo Europeo
de Desarrollo Regional (FEDER) (Spain and European Union), by grants
No.~GR21014 and No.~IB20079 (partially funded by FEDER) funded by
Junta the Extremadura (Spain), and by the Atracción de Talento program
(Ref.~2019-T1/TIC-12776) funded by Comunidad de Madrid and Universidad
Complutense de Madrid (Spain).

\bibliographystyle{ws-rv-van}

\end{document}